# BASBA: A Framework for Building Adaptable Service-Based Applications

Kavan Sedighiani,[1] Saeed Shokrollahi[2] and Fereidoon Shams,[3*]

[1,3] Department of Computer Science and Engineering, Shahid Beheshti University (SBU), Tehran, Iran
[2] Cyberspace Research Institute, Shahid Beheshti University (SBU), Tehran, Iran

[1] sedighiani@parsimap.com, [2] s_shokrollahi@sbu.ac.ir, [3] f_shams@sbu.ac.ir

**Abstract**
Due to the continuously changing environment of service-based applications (SBAs), the ability to adapt to environmental and contextual changes has become a crucial characteristic of such applications. Providing SBAs with this ability is a complex task, usually carried out in an unsystematic way and interwoven with application logic. As a result, developing and maintaining adaptive SBAs has become a costly and hardly repeatable process. The objective of this paper is to present a model-based approach to developing adaptive SBAs which separates development of adaptation concerns from development of SBAs behaviors. This approach aims to facilitate and improve the development of adaptive behaviors. In this paper, the process of developing an adaptive SBA is defined as specifying adaptive SBA models based on a metamodel and reusable adaptation tactics. These models are then transformed into runtime model artifacts and running system units performing runtime adaptive behaviors. The approach introduces a systematic method to derive adaptation behaviors from adaptation models, which facilitates the development of adaptive behaviors. The empirical evaluations in three studies show that our approach enhances the development of adaptive behaviors in terms of identifying more proper adaptation plans, reducing the development time, and increasing understandability, modifiability, and correctness of code.

## 1. Introduction

In today's dynamic environment of enterprises, collaboration among information systems has become essential to success. In this context, service-based applications (SBAs) offer promising potentials by enabling enterprises to define their business processes based on composition and coordination of software services [1-3], which may be owned by the application developer or a third party [4].

SBAs are meant to operate in a distributed, non-deterministic, unpredictable, heterogeneous, and highly dynamic environment [5]. At the same time, SBAs should be dependable in the sense that they should meet the Quality of Service (QoS) requirements [6]. These requirements highlight the need for adaptive SBAs to cope with changes and dynamics in the environment in order to demonstrate a better tradeoff among required quality attributes. However, current solutions for developing SBAs often lack proper mechanisms for modeling adaptive behaviors [7], or do not realize adaptation mechanism at implementation levels. Research on business process management (BPM) and service-oriented architecture (SOA), for example, has mainly focused on the ability to select and dynamically substitute services at runtime or at deployment time [6,8], paying less attention to the problems of how adaptation behaviors should be performed at service-collaboration level as a continuous process.

Developing and maintaining adaptive SBAs is a complex task that poses several engineering challenges [9]. In this regard, SBAs have been investigated from a variety of perspectives including introducing control flow mechanism [10], process family models [11], product line engineering [12-14], defining a collection of related process variants [15], and managing contextual properties dynamically [16-19]. However, there has been no proper framework taking into account the adaptability aspects in developing SBAs [20]. The problem with these approaches is that they do not introduce a systematic way to separate adaptation logic from adaptable system. Furthermore, they do not describe and perform adaptation and reconfiguration in a generic and reusable way. In these approaches, coordination models of services and adaptation behaviors are interwoven, increasing the complexity and reducing the maintainability of such systems. These issues make the development and maintenance of adaptive SBAs a challenging task. To overcome this challenge, the activities related to adaptation behaviors logic should be managed as an explicitly separate concern. For this aim, a proper model is needed to support the implementation and execution of the adaptation logic.

In this paper, we introduce a new approach to developing adaptive SBAs relying on the role of runtime models within feedback loops as the knowledge of adaptation. It should be emphasized that the knowledge of adaptation should come from the domain knowledge, where the SBA is developed and evolved. It should also be considered as the main logic of the adaptation by representing the knowledge of adaptation as runtime models. Therefore, we make a causal connection between runtime models, derived from design-time models, and the SBA. To this end, we present a new way to model adaptation aspects of an SBA as runtime models, aiming to provide the expressiveness for designing models required for the runtime adaptation of SBAs. This approach enables service integrators to: i) describe adaptation requirements of an SBA, and then ii) derive the right runtime adaptive behaviors, without increasing the complexity of the service development.

More concretely, the objective of the proposed approach is defined to facilitate and improve developing adaptive behaviors in SBAs in terms of development time and quality of developed adaptive behaviors. Through applying BASBA framework to two case studies and an academic environment study, we have confirmed the effectiveness of BASBA in identifying more appropriate adaptation plans and implementing more understandable, modifiable, and semantically correct adaptation behaviors in response to adaptation needs.

In addition, the results show that employing BASBA reduces the development time of adaptive behaviors.

The rest of the paper is organized as follows. Section 2 gives an overview of the related work. Section 3 shows a motivating example to illustrate the need for runtime adaptation. Section 4 introduces a framework for building adaptive SBAs. Section 5 explains how an adaptive SBA is designed and developed using BASBA. Section 6 describes elements of the BASBA framework and runtime models. In Section 7, we apply the framework in practice and discuss the results. Finally, Section 8 serves as the paper's conclusion where we present our future work directions.

## 2. Related work

Adaptation can be defined as a process of modifying an SBA in order to satisfy new requirements and to fit new situations dictated by the environment on the basis of adaptation strategies designed by the system integrator [3, 21]. Conventionally, dynamic adaptation has been managed within the application logic at the code level. For example, mechanisms such as exception handling or timeouts with some fixed hardcoded alternatives are common approaches to detect faults or system anomalies and resolve them. Although these mechanisms are usually supported by modern programming languages, their main disadvantage is that it can be difficult to develop and maintain adaptive functionalities as they would be interwoven with application logic.

Normally, service selection and binding are used as a key mechanism for adaptation in adaptive service-based applications. In this approach, adaptability is defined as a way to select the best set of services available at runtime for dynamically configuring and executing abstract workflows, taking into consideration adaptation requirements such as process constraints, user preferences and the execution context [22, 23]. Calinescu *et al.* [24] try to adapt service-based systems by dynamically adjusting service selection, resource allocation, and relevant parameters on the basis of quantitative verification and probabilistic logic to decide on the best adaptation. In the mentioned approach, the problem of adaptive service composition is stated as follows: creating and executing a workflow that satisfies the functional and non-functional requirements of the service, while being able to continually adapt to dynamic changes in the environment [25] given the specifications of a new service. The main problem with the approach is that adaptation is limited to service selection on the basis of defined abstract workflow models.

Workflow adaptation approach is another main trend in this area aiming to provide adaptability of workflow instances at build-time or at runtime by introducing structural changes of workflow elements to the atomic parts of the workflow [7]. Several systems based on workflow adaptation approach, also called agile workflow systems, have been implemented to facilitate structural changes of workflows at runtime. In this approach, workflow instances can be created and tailored to a particular need, and can be adapted according to the situation after they have been started. The two main elements of the approach are: i) configurable workflows, allowing to define workflow elements that can be switched on or off at runtime, and ii) exception handling, allowing to annotate workflows with exception handling patterns. However, most of these approaches are based on single basic changes and need an expert to guide the runtime adaptation process. In addition, they often lack the necessary instructions to become adaptable to a given context. As a result, they are inefficient in more dynamic environments in which changes have to be managed more frequently and systematically. Changing patterns [26-27] is another similar approach providing a way of modeling high-level change operations instead of specifying a set of change primitives to realize the desired adaptation model. Examples of change patterns include the insertion and deletion of process fragments, or embedding them in loops [28]. However, this approach usually provides the process designer with only those change patterns that allow transforming a sound process model into another sound one, imposing structural restrictions on process models [28]. Therefore, the approach is restricted to only a limited set of designed patterns. Similarly, many different approaches have been proposed in recent years to develop applications that can be customized at runtime using dynamic software product lines [14, 29]. However, this approach usually results in restricting the potential customizability of SBAs and is limited to a given set of changes.

Explicit management of knowledge in the adaptation cycle can remarkably enhance the benefits of runtime adaptation. Therefore, some studies have employed runtime models as the knowledge within feedback loops for adaptation [30]. Blair *et al*. in [31], define a runtime model as "a causally connected self-representation of the associated system that emphasizes the structure, behavior, or goals of the system from a problem space perspective." As a basis for self-adaptation, the use of architectural models has a number of useful properties such as providing a global perspective on the system, preserving integrity constraints, and helping to ensure the validity of any change [32].

In this regard, several approaches based on autonomic computing and self-adaptive systems engineering have been proposed to address the challenge of separating adaptation logic and adaptable systems [33,34]. Among these approaches, those in which system models, particularly software architectural models, are maintained at runtime seem more promising. A number of architectural approaches have been proposed to address the problem of managing the design complexity of self-adaptive systems [9,35-38]. At the heart of many of such adaptation techniques, there is a component capable of designing, at runtime, a strategy for adapting to the changes in the environment, system, and requirements [38]. In addition, several goal-driven approaches are introduced to model the variability and guide the architectural design based on goal-oriented requirement engineering to provide a basis for the engineering of self-adaptive systems [39-41]. Goal-driven adaptation puts the emphasis on the requirements that need to be solved by the managing system for the concrete realization of self-adaptive systems [42]. In this direction, model-driven approaches that directly execute the feedback loop via model interpretation are introduced to the development of adaptation engines [43-47]. The aim of these approaches is to support the explicit specification and execution of feedback loops. However, the main issues for developing adaptive SBAs, here, are as follows: what are the main elements of runtime models, how they should be generated and updated, and how they should be used in the development lifecycle of SBAs.

The development lifecycle of SBAs must enable the systems to be dynamic. Therefore, considering the continuous adaptation and evolution aspects of the systems are important or even essential for service-oriented development lifecycle models [4]. Various lifecycle models have been introduced to develop SBAs [34, 48]. Among these lifecycle models, S-Cube [49] is specifically intended to facilitate the adaptation of SBAs. S-Cube describes developing SBAs as two interrelated cycles: the evolution cycle and the adaptation cycle, and claims to not only SBAs go through the transition between the runtime operation and the evolution phases in order to be continuously improved, but also that they should be provided with mechanisms at runtime for the automatic detection of problems, identification of possible adaptation strategies, and enacting these strategies. Although S-Cube expands the standard development process of SBAs to consider the runtime aspect of SBAs and adaptive behaviors, it neither introduces mechanisms for developing adaptive behavior nor specifies the role of runtime models in the adaptation cycle. Considering the above-mentioned problems, there is a need for a systematic approach to develop adaptive behaviors based on specific engineering of adaptation engines and feedback loops.

To address the mentioned problems, we propose a framework to improve and facilitate developing adaptation plans. It presents a systematic way to keep adaptation concerns and behaviors at the design level separate from execution models. The proposed framework makes it possible to automatically derive right runtime adaptive behaviors from adaptation requirements, in conformity with a model-based approach. It provides mechanisms for modeling and realizing adaptive behaviors at service-collaboration level. Developing adaptive behaviors in this framework relies on reusable adaptation tactics. Therefore, it is not limited to service selection or a limited set of designed patterns. Employing reusable adaptation tactics results in performing adaptation in a generic and reusable way, reducing the complexity and increasing the maintainability of the system.

# 3. Motivating Example

To illustrate the need for runtime adaptation, we introduce a sample composite service that supports handling a simplified emergency case as an example. Fig. 1 shows this example in business process management notation. The business process starts when an emergency call is received by a call-taker. At first, the calling number and the location of the incident should be identified. Next, the call-taker needs to input additional information about the incident. After collecting of the essential information, the proper fire station(s) should be selected and asked to assign and dispatch field personnel. During the mission, the position of the personnel should be sent to management center, where they are monitored and their location is displayed on a map.

Although the above example is an oversimplified scenario, there are quite a few possible environmental and contextual changes and events. For example, the caller Id may not be identified or the service finding geographical location based on caller Id may be unavailable or response too late. All these events need to be handled at runtime if they occur. Otherwise, the process may lead to unsatisfactory situations. Therefore, proper compensation strategies should be provided and made available at runtime for any situation.

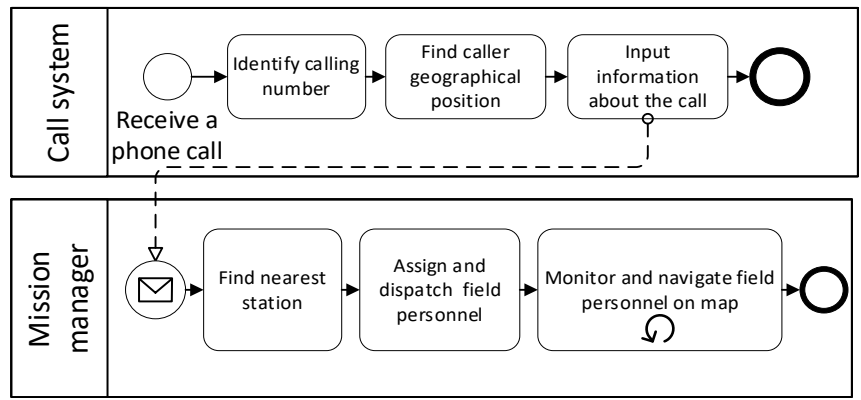

**Fig. 1** A simplified process model of an emergency call system

# 4. BASBA framework

In this section, we introduce a novel approach to facilitate building adaptive SBAs called BASBA (Building Adaptable Service-Based Applications) framework, which offers adaptive behaviors in composition and coordination of services. We emphasize that the logic of adaptation behavior should be determined by design-time models that are then transformed into runtime models. The BASBA framework facilitates designing and developing adaptive behaviors. For this purpose, we extended the S-Cube lifecycle [49] to consider runtime models (Fig. 2). Like S-Cube, the BASBA framework describes developing SBAs as two interrelated cycles: the evolution cycle and the adaptation cycle. In the evolution cycle shown on the right-hand side of the figure, the service integrator concentrates on the development of the adaptive SBA and defining the required quality attributes. However, the adaptation cycle considers providing mechanisms for adapting the system at runtime through automatic detection of problems, identification of possible adaptation strategies, and enacting these strategies.

As shown in Fig. 2, in the evolution cycle, through the "specifying and designing adaptive SBA" phase, the service integrator develops a set of design models. These design models describe the coordination of services, the required qualities, and adaptation behaviors. In this phase, BASBA provides a BASBA notation and a set of reusable adaptation tactics. The BASBA notation describes how an adaptive SBA should be specified. It includes elements to model adaptation plans and QoS requirements. Adaptation tactics determine how a process should be changed. The design models, in the "generating adaptive process instances" phase, are transformed into runtime coordination services. The coordination services are deployed on the process executor and aggregator engine to form the target system. Through the "operation and management" phase, the service integrator gets feedback to update the design models. The design models are also used to generate adaptive runtime models to monitor and adapt the system in the adaptation cycle.

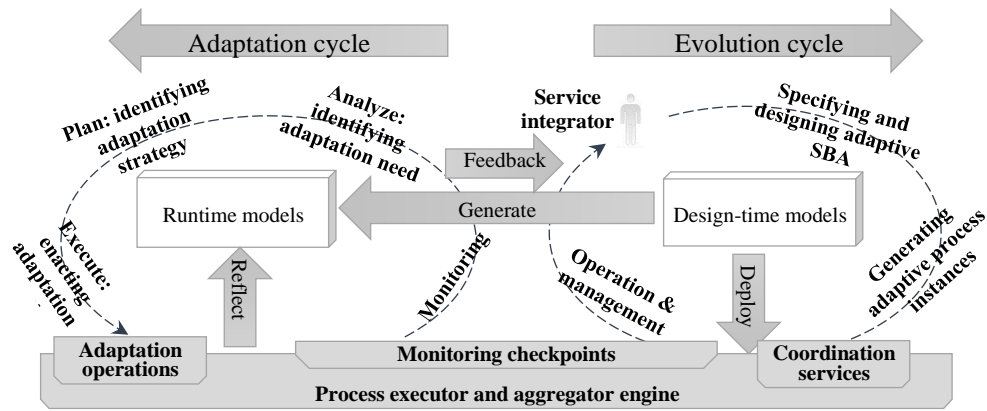

**Fig. 2** The role of models in the lifecycle of an adaptive service-based application

To manage adaptive behaviors in BASBA, the source, the elements, and the role of runtime adaptation models should be identified. To this end, we define an adaptive SBA in three layers: the specification layer, the runtime adaptation layer, and the execution layer (Fig. 3). These layers are inspired by the three-layer architecture introduced by [35]. In BASBA, the logic of each layer is determined by the higher layer.

The top layer includes process models, their possible variations, and the required QoS objectives designed by a service integrator through the evolution cycle. These models are known as the source of adaptation, which are transformed into runtime models. The runtime models determine the adaptation logic of the system including the way the system is monitored and adapted at runtime, which form the runtime adaptation layer. This layer takes the control of the executing system at specified checkpoints and, based on the adaptation needs, changes the behavior of the system using adaptation tactics.

The bottom layer is the execution layer, which consists of monitoring interceptors, change actions (actuators), and runtime processes managed by process executor and aggregator. The bottom layer consists of a set of runtime services connected by connectors that accomplish the purpose of the system. At this layer, interceptors, which are placed on connectors, facilitate reporting the measured data to higher layers.

Runtime models have a significant role in the BASBA adaptation cycle for runtime adaptation. These models can significantly separate the concern of adaptation at design level from the execution level, and facilitate the development of adaptive behaviors. At this layer, the role and the elements of runtime adaptation models are determined. Following the MAPE-K (Monitor, Analyze, Plan, Execute, and Knowledge) control loop [50], the BASBA framework consists of a control loop which is periodically executed.

Monitoring models determine the location of checkpoints and the granularity of the collected sample data. The collected data is used to update the service execution models and manage SBA application through an adaptation cycle, or to start another evolution cycle. Runtime models (service composition and coordination models) are analyzed on the basis of evaluation models such as constraints, time, and failures to identify adaptation needs.

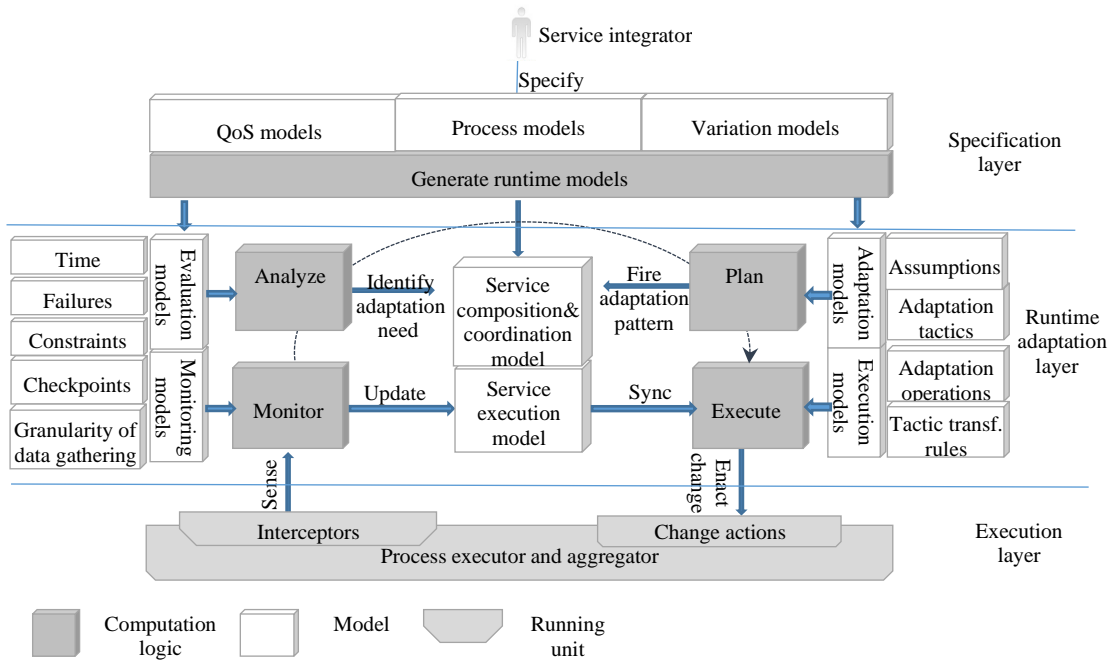

**Fig. 3** The three layers of adaptability in BASBA

When an adaptation need is identified, an adaptation pattern is fired in response to the informed adaptation need. The Adaptation pattern is a realization of a variation model defined as adaptation plan by the service integrator. It includes assumptions, a flow of adaptation tactics, and the expected results. Finally, the adaptation plan is transformed into the executing system to form a new variation of coordination services. The transformation is done through adaptation operations and transformation rules that are both defined by adaptation tactics.

For example, in developing adaptive behavior in the motivating example, the service integrator may have concerns about network bandwidth, which may increase communication delay between the management center and the field personnel. The service integrator must define the required QoS objectives and the possible variations as the source of adaptation. The QoS objective can be an acceptable response time between the management center and the field personnel. The service integrator can define this objective by fuzzy linguistic words such as "less than 1 second is acceptable, between 1 to 3 seconds is tolerable, and more than 3 seconds is inacceptable". In response to this objective, the service integrator can design a plan (called an adaptation pattern at BASBA level) to reduce the volume of data by using a tactic such as compressor and decompressor. This decision is called a *variation* in BASBA.

In this case, two checkpoints can be located at the services that send and receive messages between the management center and the field personnel. The data gathered in these checkpoints forms a model which determines the average response time. The evaluation model, in this case, can be defined as consisting of two rules:

i) if the average response time is *inacceptable*, then fire a *hard* adaptation need to decrease volume of data.

ii) ii) if the average response time is *tolerable*, then fire a *soft* adaptation need to decrease volume of data.

These rules are slightly different. Both rules fire a need for adaptation to decrease volume of data, however, the first rule fires a hard need and the second one fires a soft need. An adaptation need can fire an adaptation pattern. The adaptation pattern in this case can be defined as follows:

i) if battery is *high*, and a *soft* adaptation need is fired to *decrease the volume of data*, then execute the *compressor/decompressor tactic between the services that send and receive messages.*

ii) ii) if battery is *high or medium*, and a *hard* adaptation need is fired to *decrease the volume of data*, then execute the *compressor/decompressor tactic between the services that send and receive messages.*

Both adaptation patters in this case have the same consequences, however, each one can be fired in different conditions.

# 5. Designing adaptable SBAs using BASBA

In this section, we describe how adaptive behaviors for an SBA can be designed through "specifying and designing an adaptive SBA". In BASBA framework, an adaptive behavior is defined by adaptive process models. An adaptive process model is an assembly of the workflow logic of the process (process model), the QoS model of the SBA (quality requirements), and the possible variations (adaptation plans), which can be converted to BASBA components to form the runtime models and the running system.

In the following parts of this section, first, we explain the adaptive process model and the related elements through the motivating example. Next, in Section 5.2, we introduce the formal definition of the adaptive process model on the basis of process algebra and first-order logic. We describe how quality requirements (QoS model) and adaptation plans (variation models) should be defined for a process model to form an adaptive process model. Then, in Section 5.3 and Section 5.4, we explain the QoS model and variation models in more detail. In Section 5.4, we also introduce the BASBA notation and elements to define the adaptive process including QoS components and adaptation tactics. The BASBA notation facilitates defining the adaptive process model.

### 5.1. Explanation of the adaptive process model through the motivating example

In order to define adaptive behaviors for the motivating example, the workflow logic and the related quality requirements and variation models should be specified. The workflow logic can be modeled in an abstract way, using Business Process Management Notation 2.0 (BPMN). For example, in Fig. 1, the BPMN notation is used to model the workflow logic of a process in the emergency call system. Activities in the workflow represent abstract services, which should be mapped to concrete services that satisfy them. For example, "find caller geographical position" is an abstract service that can be satisfied by "find caller geographical position by Id" or "find caller geographical position on map".

When an abstract business process is defined, the service integrator defines the QoS model. The QoS model plays a crucial role in triggering adaptation plans and analyzing the tradeoff among quality attributes. BASBA can consider different measurable QoS attributes to specify an SBA called *quality requirements*. Quality requirements can be defined on the basis of some measurable properties such as time, data value, failure, or a mathematical function. For example, the response time of "identify call number" service can be defined as a quality requirement. For each quality requirement, the service integrator, defines a fuzzy measure using triangular fuzzy linguistic words. For example, the fuzzy measure for response time quality requirement in the "finding geographical location" process part can be shown as (-, 10 seconds, 30 seconds, per instance), which means the response time for execution of that part of the workflow is totally acceptable if it is less than 10 seconds, the response time between 10 to 30 seconds is tolerable, and the response time over 30 seconds is inacceptable and triggers an adaptation plan. Another example can be the availability of the map service defined on the basis of failure rate, which is the ratio of failures to the total number of the process calls. The fuzzy measure for this can be defined as (+, 0.96, 0.99, monthly), which means if the availability of the service in one month is inacceptable to be less than 0.96, the availability between 0.96 and 0.99 is tolerable, and the availability over 0.99 is totally acceptable. Each fuzzy measure can fire an adaptation trigger. For example, when the response time for "identify call number" is not acceptable, the "automatic call number detection failed" adaptation trigger should be fired.

When the QoS model is defined, the service integrator specifies the variability model of the workflow. Table 1 shows some possible adaptation plans for the emergency call system introduced in the motivating example. For the sake of simplicity, we have used the natural language to define the adaptation plans and tactics. As shown in Table 1, an adaptation plan is started when an adaptation trigger occurs. For example, the adaptation plan #1 is triggered when an automatic call number detection fails. For each adaptation plan, some specific adaptation tactics are defined on the basis of reusable adaptation tactics. Each option is obtained by specifying a generic adaptation tactic. However, only the options whose pre-assumption can be verified by the context can be applied. For example, in the adaptation plan #2 replacing "find caller geographical position by Id" with "find caller geographical position on map" can only be applied when the "map service is available" pre-assumption could be verified by the context. Each adaptation plan should be specified using the BASBA adaptation plan notation.

**Table 1** Some possible adaptation plans for the emergency call system

| Plan Id | Adaptation trigger | Reusable adaptation tactic | Specific adaptation tactic | Pre-assumption | False assumption |
|---|---|---|---|---|---|
| #1 | Automatic call number detection failed | Process variation – replace activity | O1: Replace (automatic call number detection, input call number manually) | Human operator is available | Soft: Increase response time |
| | | Process variation – skip activity | O2: Skip (automatic call number detection) | - | Hard: Id is identified |
| #2 | Falsify: Id is identified | Process variation – replace activity | O1: Replace (find caller geographical position by Id, find caller geographical position on map) | Map service is available | Soft: Increase response time |
| | | Process variation – replace activity | O1: Replace (find caller geographical position by Id, input caller text address) | Human operator is available | Hard: Caller location is identified |
| #3 | Map is unreliable | Activity variation - serial execution | O1: Serial execution (municipality map, google map) | Google map is available | Soft: Increase response time |
| | | Activity variation – parallel execution | O2: Parallel execution (municipality map, google map) | Google map is available | Soft: Increase resource utilization in google map service |
| #4 | Network is unreliable | Communication variation- add queue | Add queue (sending message to vehicles) | - | Soft: Increase memory utilization on client device |

| #5 | Low communication bandwidth | Communication variation- compressor and decompressor | Add compressor (send vehicle data), add decompressor (receive vehicle data) | - | Soft: Increase in battery utilization |

Each adaptation plan can cause false assumptions, resulting in another adaptation trigger. There are two categories of false falsifications: *hard falsifications* and *soft falsifications*. Hard falsifications can have a chain effect, causing an event to trigger another adaptation plan. Soft falsifications may result in undesirable quality requirement requiring tradeoff analysis. Therefore, before enacting an adaptation plan in response to a soft falsification, a tradeoff should be made between enhancements that can be obtained from enacting the plan and its negative effect on other quality attributes based on defined QoS models. In the case of adaptation plan #1, there are two options: "replace automatic call number detection with input call number manually" and "skip automatic call number detection". The former plan can be applied when the "human operator is available" pre-assumption can be verified by the context. However, both plans may result in false assumptions. The first plan increases response time, which is a soft falsification, and the second one will falsify "Id is identified", which is a hard falsification and necessitates executing another adaptation plan.

Fig. 4 shows an example of an adaptation plan for the motivating example. This plan means that there should be an exception evaluation unit in "identifying the calling number" to detect occurrence of a failure, which may raise the "Id is identified" false assumption, resulting in executing the adaptation plan. In this plan, at first, the skip tactic is enacted to remove the "find caller geographical position by Id" activity from the process. Then, there are two alternatives. First, the "find caller geographical position on map" service can be added, provided the assumption "map service is available" is held. Second, the "input address in text format" service can be added, but this will lead to another false assumption that might result in executing another adaptation plan.

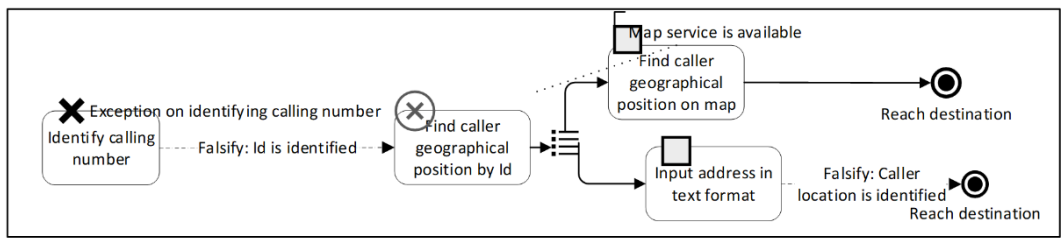

**Fig. 4** An example of an adaptation plan in BASBA notation

## 5.2. Adaptive process model

The specification layer defines an adaptation process model. An adaptative process model is an assembly of activities connected by flow objects enriched with QoS requirements, a set of assumption to be held and a set of adaptation plans. QoS requirements are defined on the basis of measurable properties. An adaptation plan is a representation of an adaptation pattern in BASBA adaptation plan notation introduced in the following section. In BASBA, each adaptive process model, called $APM$, is an aggregation of $W$, $QR^+$ and $AP^+$ and takes the form:

$$APM ::= (W, QR^+, AP^+) \qquad (1)$$

Where $W$ is a workflow, $QR^+$ is a set of quality requirements that define the QoS model, and $AP^+$ is a set of adaptation plans that define the variation model.

The workflow is defined using process algebra in which each process, called $P$, takes the form:

$$P ::= Seq(P^+) \mid Loop(P) \mid Sel(P^+) \mid and\_par(P^+) \mid Opt(P) \mid S \qquad (2)$$

Where $Seq(P^+)$ is a sequence of processes, $Loop(P)$ is a loop on a process, $Sel(P^+)$ is a selection of processes, $and\_par(P^+)$ is an and-parallel of processes, $Opt(P)$ is an optional of a process, and $S$ is a service, defined by a service specification and can be delivered by a set of service providers.

The QoS model is defined by a set of quality requirements ($QR^+$). Each $QR$ is defined on a process $P$, and evaluated by measurable property $MP$ and takes the form:

$$QR ::= (P, MP, FM, TR) \qquad (3)$$

Where $P$ is the part of the process which is evaluated, $MP$ is a measurable property that is measured in $P$, $FM$ is a fuzzy measure that is used to evaluate the $MP$, and $TR$ is an adaptation trigger, which will be triggered if the fuzzy measure shows a violation. The measurable properties are explained in next section, which takes the form:

$$MP ::= \text{Time} \mid \text{Data} \mid \text{Failure} \mid \text{Count} \mid \text{Constraint} \mid \text{Derived} \mid \text{Aggregated}$$

Each fuzzy measure is defined using a triangular fuzzy term which defines the satisfaction level of MP, and takes the form:

$$FM ::= (O, X_1, X_2, TI) \qquad (4)$$

where $O$ can be '+' or '-', with '+' indicating the greater value the better result, and '–' indicating the lower value the better result, $X_1$ and $X_2$ forms three triangular fuzzy terms, and and $TI$ indicates the time interval to calculate the $MP$ in average. For example, if $O$ is '-', it means $QR$ value less than $X_1$ is totally acceptable, the value between $X_1$ and $X_2$ is tolerable, and $QR$ greater than $X_2$ is not acceptable.

The variation model is defined based on adaptation plans. Each adaptation plan, called $AP$, is an aggregation of $TR$, $AT^+$, $PA^+$, and $FA^+$ and takes the form:

$$AP ::= (TR, AT^+, PA^+, FA^+) \qquad (5)$$

Where $TR$ is an adaptation trigger that can be fired from a $FM$ or can be the result of executing an $AP$, $AT^+$ is a set of adaptation tactics, $PA^+$ is a set of pre-assumptions that should be satisfied to run the adaptation plan. $FA^+$ is a set of false assumptions that is the consequences of running the adaptation plan. Each *FA* can trigger a *TR* result in firing another adaptation plan. Adaptation tactics are introduced in Section 6.

### 5.3. The QoS model

In order to specify and design QoS models, a service integrator defines quality requirements on the basis of some predefined measurable properties. The measurable properties in BASBA are modeled on the basis of process performance indicators introduced in PPINOT ontology [51]. These measurable properties can be a single-instance measure such as time, failure, count, data, constraint or derived, or can be a multi-instance measure such as derived or aggregated. To calculate a multi-instance measure, a set of process instances are used. The availability of the map service in a period of time (described in Section 5.1) is an example of multi-instance measure implemented by an aggregate function. The description of measurable properties is explained in Table 2.

**Table 2** The description of measurable properties used in BASBA

| Measure type | Description |
| --- | --- |
| Time | It measures the duration of time between two time instants |
| Failure | It is a Boolean value that indicates a failure |
| Count | It measures the number of times something happens |
| Data | It measures the value of a certain part of a data object |
| Constraint | It is a Boolean value that measures the fulfilment of certain condition on process instances |
| Derived | It is defined as a mathematical function over one or more measure definitions. There are two types of derived measures depending on whether the measure definitions are single-instance or multi-instance measures |
| Aggregated | It is defined by aggregating one of the previous measures in several process instances using an aggregation function such as sum or average |

In BASBA, a model is developed for each measurable property to define the way the data should be collected and analyzed. For each model, some transformation rules have been developed to generate required interceptors. Interceptors are units that gather data at runtime to form the execution profile and update the execution model of the process. The execution models are analyzed on the basis of evaluation models. An evaluation model is defined as a function of measurable properties to identify an adaptation need called a false assumption. In evaluation models, the acceptable range of each metric is defined using triangular fuzzy linguistic words showed in Equation 4. For constraint and failure measurable properties, the fuzzy measure is defined using count or derived functions on these properties.

The service integrator should consider the structure of processes to calculate measure properties. In this regard, the service integrator can implement its own derived and aggregate functions, using delegate functions, to measure different quality requirements. In BASBA, some derived measures that use the functions per structure of processes and QoS attributes measures, introduced in [52], are implemented. Table 3 shows the implemented derived measurable properties in the BASBA framework.

Table 3 Functions per structure of process and QoS attribute

| Structure of process | Sequence | Loop | Selection | Parallel |
|---|---|---|---|---|
| Response Time(T) | $\sum_{i=1}^{n} T(S_i)$ | $k * T(S_i)$ | $\sum_{i=1}^{m} P_i T(S_i)$ | $\max\{T(S_i)_{i \in \{1..p\}}\}$ |
| Cost(C) | $\sum_{i=1}^{n} C(S_i)$ | $k * C(S_i)$ | $\sum_{i=1}^{m} P_i C(S_i)$ | $\sum_{i=1}^{p} C(S_i)$ |
| Availability (A) | $\prod_{i=1}^{n} A(S_i)$ | $A(S_i)^k$ | $\sum_{i=1}^{m} P_i A(S_i)$ | $\prod_{i=1}^{p} A(S_i)$ |
| Reliability (R) | $\prod_{i=1}^{n} R(S_i)$ | $R(S_i)^k$ | $\sum_{i=1}^{m} P_i R(S_i)$ | $\prod_{i=1}^{p} R(S_i)$ |

$S_i$ is a service component
$P_i$ is the probability of selecting service component $S_i$
$n$ is the number of sequential service components
$k$ is the number of iterations on a service component
$m$ is the number of service components associated with a logical condition
$p$ is the number of service components executed concurrently

### 5.4. Variation models

Variation models are another main specification of an adaptive SBA defined as adaptation plans. Each adaptation plan is a sequence of actions in response to a false assumption and determines how the process instance should be modified. Adaptation plans are designed on the basis of BASBA adaptation plan notation. Each adaptation plan starts with an evaluation unit determining when an adaptation plan should be executed, a set of adaptation tactics determining how the process should be changed, and a set of flow objects determining the sequence of execution of adaptation tactics.

In BASBA, a notation is introduced which enables the service integrator to design adaptation plans. For each notation, a template class is developed, which defines the behavior of the adaptation element at runtime and the corresponding transformation rules. The notation consists of three types: evaluation units, flow objects, and adaptation tactics. Fig. 5 shows evaluation unit and adaptation plan flow objects notations. An evaluation unit is defined on a service or a part of a process to assess a quality requirement. For this purpose, the service integrator defines a measurable property that should be evaluated on the basis of a defined fuzzy measure. BASBA supports three types of flow objects: Alternative flow, Adaptation trigger flow, and Simple flow. Flow objects determine the order of execution of the adaptation plan.

BASBA also supports 10 types of adaptation tactics. Adaptation tactics are the main elements of adaptation plan notation. An adaptation tactic is a design decision that affects the system response to some stimuli. Service integrators can reuse adaptation tactics to define adaptation plans. It should be mentioned that there is no consideration of tradeoff in tactics, and the only focus of an adaptation tactic is on a single quality response. For example, the focus of serial execution tactic is on reliability, and the focus of compressor/decompressor tactic is on network performance quality. This property allows adaptation tactics to be reusable elements in defining adaptation plans.

| Type | Title | Graphical representation | Description |
|---|---|---|---|
| Evaluation unit | Quality requirement | 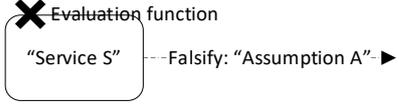 | Evaluate a measurable property in a service or a part of a process on the basis of the defined fuzzy measure. If it fails, an adaptation trigger will be fired, resulting in falsify an assumption. |
| Flow objects | Alternative flow | 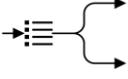 | There are two variations. If the pre assumptions of the first one does not hold, then try the next variation. |
| Flow objects | Adaptation trigger flow | -Falsify: "Assumption A"- ▶ | An adaptation trigger that falsified an assumption "Assumption A" has been fired. |
| Flow objects | Simple flow | A ⟶ B | Run "B" after "A" |

**Fig. 5** The evaluation unit and adaptation plan flow objects notations in BASBA

In Fig. 6, we have classified BASBA adaptation tactics in three groups: those acting on the structure of the workflow, those affecting the way an activity runs, and those changing the communication protocols between services. To address the first group, a variation is defined using skip, add, and replace activities in a workflow. The second category of variations refers to possible variations inside an activity which do not affect the workflow. These types of adaptation include parallel execution, serial execution, and re-execution of an activity. The third category of adaptation tactics includes those that change the way services communicate in an SBA. An example of these tactics is putting compressor/decompressor between two services in order to reduce the volume of data communication. In the motivating example, applying the compressor/decompressor tactic on the "monitor and navigate field personnel on map" process block reduces the volume of data transferred between services in the process block, results in reducing the latency of data communication. However, it also results in higher battery usage on client devices. Cache element, reducing size, and aggregating data are among other examples of communication tactics. Adaptation tactics in BASBA can be extended. To this end, the model transformation, the logic of the tactic, its effect on quality attributes, and required connectors should be developed. Section 6.3 introduces the elements of the adaptation tactics in more detail.

| Category | Title | Graphical representation | Supporting connectors | Description |
|---|---|---|---|---|
| Process variations | Skip activity | 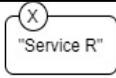 | SimpleConnector | Skip "Service R" |
| | Add activity | 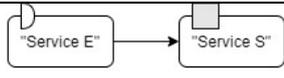 | SimpleConnector | Add "Service S" after "Service E" |
| | Replace activity | 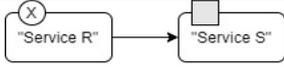 | SimpleConnector | Replace "Service R" with "Service S" |
| Activity variations | Parallel execution | 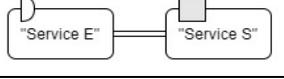 | ParallelOutConnector ParallelInConnector | Add "Service S" and execute in parallel with "Service E" and take the first response and continue |
| | Serial execution | 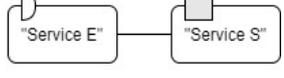 | SerialOutConnector SerialInConnector | Add "Service S" and execute if "Service E" failed, otherwise skip "Service S" |
| | Re-execution | 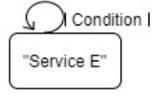 | ConditionConnector | Re-execute "Service E" until a condition "condition function" is reached |
| Communication variations | Compressor/ decompressor | 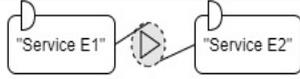 | CompressorOutConnector CompressorInConnector | Compress output of "Service E1" and decompress it before "Service E2" |
| | Aggregate data | 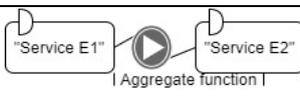 | DataModifierOutConnector DataModifierInConnector | Aggregate output of "Service E1" using "aggregate function" and disaggregate it before "Service E2" |
| | Reduce size | 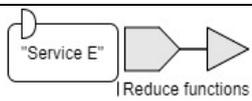 | DataModifierOutConnector | Reduce the size of output "Service E" using "reduce function" |
| | Cache element | 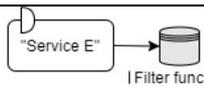 | CacheElementConnector | Add local cache element to "Service E" and cache the data based on "filter function" |

**Fig. 6** Adaptation tactics currently supported in BASBA

## 6. BASBA Runtime models

When an adaptive process model is designed, then, the design elements are transformed into runtime elements forming the runtime adaptation models. BASBA uses a metamodel to support development of adaptive SBAs. The metamodel determines the elements and the structure of the adaptive SBA at design-time and runtime. It also clarifies the relationship and the transformation logic between design-time and runtime artifacts. In this section, first, we introduce the BASBA metamodel, then we demonstrate how runtime artifacts are generated from design-time artifacts.

### 6.1. The BASBA Metamodel

BASBA metamodel provides a set of concepts to model the monitoring and adaptation requirements at the service coordination level, and transform the models into runtime adaptation logic. To this end, the monitoring and adaptation elements and the way they are assembled in a workflow are modeled. Moreover, architectural and quality-related information for adaptive SBAs and the relationships among the provided concepts are determined.

Fig. 7 outlines the metamodel and the relationships among the provided concepts in an abstract way. The BASBA metamodel consists of three layers to support developing adaptive SBAs: i) the specification layer, the top block, in which the structure and objective of the SBA are represented; ii) the runtime adaptation layer, the middle block, in which the system elements are linked to system objectives using adaptive elements; and iii) the execution layer, the bottom block, in which the elements and the state of system are reflected. Each block defines the concepts related to the corresponding layer introduced in the previous section.

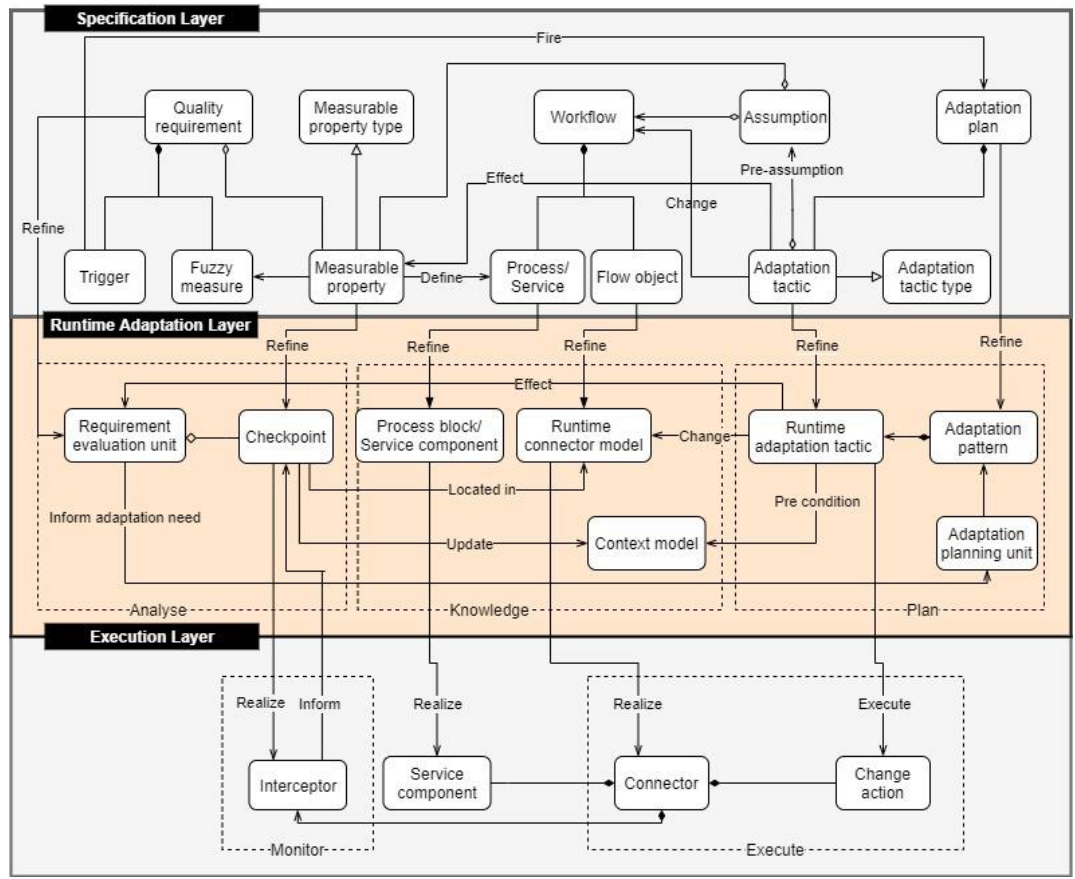

**Fig. 7** The BASBA metamodel

The specification layer in the metamodel determines the structure and relationships of the produced artifacts in design time. An adaptive process model is a workflow, enriched by quality requirements and adaptation plans. Each workflow is modeled by a set of services and processes, connected by flow objects. Quality requirements are defined on the basis of measurable properties, fuzzy measures, and adaptation triggers. Each measurable property is defined to measure a service or part of the workflow based on measurable property types. Measurable properties also form system assumptions. Each assumption is some information about the system, i. e., the estimated data based on measurable properties, or a piece of information about services, process, or flow objects in the workflow.

An adaptation plan is a set of adaptation tactics that is defined in response to an adaptation trigger. Each adaptation plan can be executed if the pre-assumptions of all corresponding tactics hold in the system. Executing an adaptation tactic changes the structure of the workflow, influences the measurable properties, results in changing the satisfaction level of quality requirements. Each adaptation tactic is defined on a part of process based on adaptation tactic types.

The runtime adaptation layer consists of models and components to control and adapt the system at runtime. These elements are generated from specification artifacts. In this layer, the structure of workflow is modeled by service components and connectors. A service component is defined as a simple service, or a process block that is a set of service components and connectors, with a start and an end. Service components can only connect to each other through connectors. Connectors determine the flow of the process and the order of execution of service components to form the process model. For this aim, each connector model has a binding model, which shows its connection to the other services and connectors.

Checkpoints are runtime units located in the runtime connector models to measure the specified measurable properties. In fact, all measurements are done in the connectors through some predefined mechanisms. The measured properties are evaluated in evaluation unites. Each evaluation unit has a fuzzy measure property to detect adaptation needs and inform the adaptation planning component.

The adaptation planning unit is a runtime component that tries to rectify the system by executing adaptation patterns. Each adaptation pattern is a realization of an adaptation plan that determines a set of adaptation tactics to be executed. An adaptation tactic is a rule-based reconfiguration of connectors based on a set of change actions defined in adaptation plan templates with predictable effects on the system context model, resulting in change in the quality requirement satisfaction level. Each adaptation tactic can be executed if the required preconditions hold in the context model. The role of the adaptation planning unit is to find the adaptation pattern(s) with the best outcome according to the quality requirements that can be executed in accordance with the current context model to change. The runtime adaptation layer changes the behavior of the executing system through changing the connectors in the execution layer. The execution layer is a set of components connected by connectors. Connectors are the main elements in forming communication between components and gathering data to analyze the system. This allows for handling the adaptation logic separately and dynamically in the communication layer, since the service components are not directly affected by changes.

In BASBA, a set of connectors and interceptors are implemented. Connectors can be added, removed, or bound to service components by change actions dynamically. The change actions provide the ability to change the connectors and the communication model at runtime, and the ability to realize adaptation tactics. In addition, a set of connectors are implemented to support simple communication and adaptation tactics (Fig. 6). For example, the parallel execution tactic is supported by ParallelOutConnector and ParallelInConnector connectors. On each connector, a set of interceptors can be installed/uninstalled at runtime to return the value of properties to the checkpoints. Moreover, a set of interceptors are implemented to support the basic measurable properties introduced in Table 2.

Change actions in the execution layer are enacted by the configuration manager component. The configuration manager is a runtime container for the deployment and modifying connectors, which provides a procedural interface for the loading, unloading and modifying connectors. The configuration manger is used by elements in the adaptation layer to realize a checkpoint by placing an interceptor on a connector, or by executing change actions determined by adaptation plans.

The BASBA framework supports SOAP and RESTful protocols for binding connectors to service components. It also supports JSON, GeoJSON, and XML data contracts. Several abstractions are defined in BASBA to implement new connectors and extend adaptation tactics. Each connector should provide a reconfiguration interface with operations such as bind and unbind. The configuration manager uses this interface to enact the change actions.

### 6.2. Generating runtime models

In this section, we describe how runtime models are transformed from specifications of adaptive process models in BASBA framework. An adaptive process model, explained in Section 5.2, includes three categories of specification models: i) workflow models, ii) quality requirements, and iii) adaptation plans. As shown in the BASBA metamodel (Fig. 7), each element of the specification models will be refined to some elements in the runtime adaptation layer and some elements in the execution layer.

A workflow in the specification layer is defined based on simple services, flow objects and process elements, which are transformed into runtime artifacts. A simple service is transformed into a service component model at the runtime adaptation layer, reflecting the concrete service that realizes the corresponding service in the workflow model. For each flow object, there is a direct transformation model that transforms the flow object into a specific runtime connector model, and a specific runtime component at the execution layer. For example, the flow object Seq (S1, S2) is transformed into a sequence connector model at runtime (including SeqInCon and SeqOutCon elements), which shows that the output of the service S1 is bound to SeqInCon, and the output of SeqOutCon is bound to service S2. The sequence connector model also maintains information about the service components (such as SC1 And SC2) that realize the S1 and S2 services. At the execution layer, the sequence connector model is transformed into the runtime connector components that connect the outbound of the service component SC1 to the inbound of the service component SC2. Fig. 8 shows the flow objects model at runtime in BASBA. Each process in the specification layer is transformed into a process block at the runtime adaptation layer, which is defined by a start connector and an end connector. These connectors provide the ability to detect when the process block is started and when it is finished.

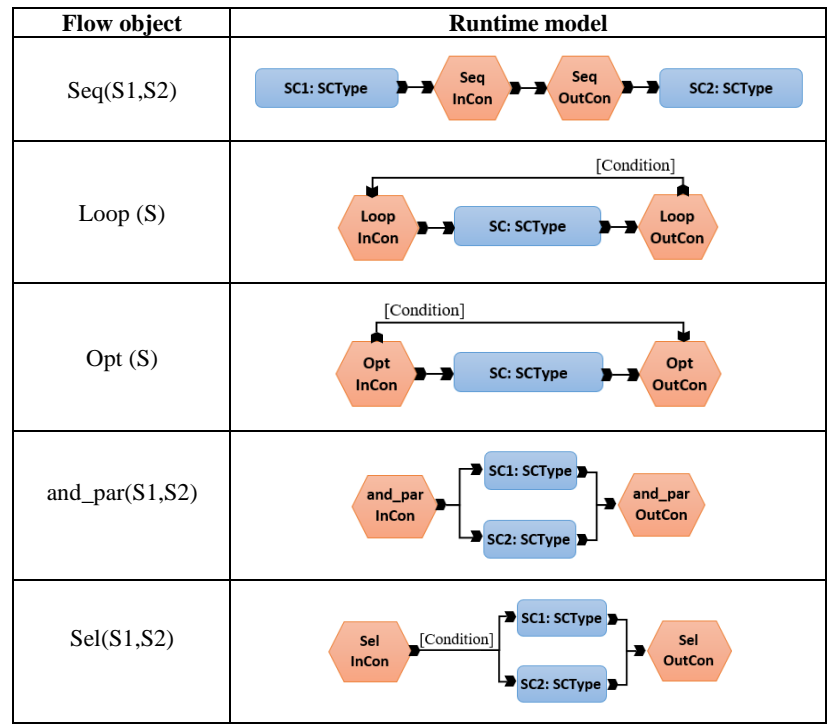

**Fig. 8** The runtime models of process flow objects

A quality requirement in the specification layer is defined on a process with a measurable property, a fuzzy measure, and an adaptation trigger. For each measurable property, a transformation rule is developed that determines how it is transformed to runtime components and how data should be collected in connectors. Table 4 shows the transformation logic of measurable properties in an abstract way. A measurable property is a basic property, or a function that is defined on the basis of basic properties. The function can be a predefined function in BASBA (Table 3), or can be a delegate function. The transformation of measurable properties into runtime elements is supported by the embedded mechanisms designed in BASBA connectors allowing to install/uninstall interceptor functions. The value of a measurable property is calculated by the checkpoints. Checkpoints are runtime components that determine the placement of interceptors on the connectors and the data that should be collected. The data gathered by the checkpoints is evaluated in the requirement evaluation units that realize the logic of fuzzy measures. Each evaluation unit is a runtime component that implements a fuzzy measure model on the measurable property to detect violations and fire an adaptation trigger. For example, specification of a quality requirement in the motivating example can be defined as follows.

$QR ::= (\text{"Finding geographical location"}, RT: Time, (Per\ Call, 2\ ms, 10\ ms), \text{"Service is Slow"})$

Where RT is a time measurable property defined on the "finding geographical location" process block. The measurable property is transformed into a checkpoint, which manages the placement of an interceptor on the connector before, and an interceptor on a connector after, the process block. The checkpoint implements functions to receive the data from these interceptors and calculates the average response time. The quality requirement $QR$ is also transformed into an evaluation unit, which implements the $(Per\ Call, 2\ ms, 10\ ms)$ fuzzy measure logic to fire the "$Service\ is\ Slow$" trigger, based on the received data from the corresponding checkpoint and the fuzzy measure function.

The data gathered at the checkpoints are also used to form the system context model. The context model is the state of the system at runtime, known as assumption in design time. In the context model, in addition to the measured data, the state of all service components, connectors, and bindings are maintained.

**Table 4** The transformation logic of measurable properties

| Measure type | Transformation rule |
|---|---|
| Time | When a time measure property is defined on a process block, the property is transformed to two interceptors on the connectors located before and after the process block, triggering time events. The time measure property is also transformed to a checkpoint object in the adaptation layer, receiving time events and calculating the execution time of the process block and updates the context model. |
| Failure | When a failure measure property is defined on a process block, the property is transformed to two interceptors on the connectors located before and after the process block, and a container to run the process block. The failure measure property is also transformed to a checkpoint object in the adaptation layer, receiving the events of starting the process block, completion of the process block, and a failure event by the container if a failure accrues during the execution of the process block. |
| Count | When a count measure property is defined on a process block, the property is transformed to an interceptor on the connector located after the process block. The count measure property is also transformed to a checkpoint object in the adaptation layer, receiving the number of times the process block was executed. |
| Data | When a data measure property is defined on a process block, the property is transformed to an interceptor on the connector located after the process block. The interceptor measures the value of data based on the process output and sends an event to the corresponding checkpoint. |
| Constraint | A constraint measure property is defined on a process block with a delegate function defining a condition. The constraint measure property is transformed to an interceptor on the connector located after the process block. The interceptor evaluates the condition based on the process output and sends an event to the corresponding checkpoint. |

| Derived | A derived measure property is defined as a mathematical function based on basic measurable properties. The mathematical function is transformed to a checkpoint component at runtime, which calculates a value based on the basic measurable properties. |
|---|---|
| Aggregated | An aggregate measure property is defined as an aggregation function based on basic measurable properties. The property is transformed to a checkpoint component to aggregate the value of basic measurable properties at runtime. |

An Adaptation plan in the specification layer is defined based on the adaptation plan flow objects (Fig. 5) and a set of adaptation tactics to manipulate a process at runtime. The adaptation plan flow objects are transformed into an adaptation pattern in response to a specified trigger, which handles the execution of runtime adaptation tactics according to a set of assumptions. Each adaptation tactic is transformed into a runtime adaptation tactic on the basis of an adaptation template class. A template class is defined by the supporting connectors (Fig. 6), precondition, pre-state, change action, post-state, and the expected effect that can instantiate a concrete adaptation tactic at runtime. A concrete adaptation tactic is instantiated through a template based on the parameters defined by the adaptation plan specification to manipulate the execution layer through connectors. The elements of adaptation plans are not directly transformed into artifacts at the execution layer, but they change connectors and bindings in the execution layer in accord with the defined change actions. In the next section, adaptation patterns and adaptation tactic templates are explained in more detail.

### 6.3. Adaptation at runtime

In BASBA, the logic of adaptation is defined according to the information held in the context model. The information includes data about system components and connectors, measured properties and requirement satisfaction levels. The data is maintained in simple propositions shown in Table 5.

**Table 5** Proposition syntax in context model

| Proposition syntax | Proposition description |
|---|---|
| $SC: ServiceComponent$ | SC is a service component |
| $SC: SCType$ | Type of the service component SC is SCType |
| $Con: Connector$ | Con is a connector |
| $Con: ConType$ | Type of connector Con is ConType |
| $bind(SC, Con)$ | SC is bound to Con |
| $bind(Con, SC)$ | Con is bound to SC |
| $MP = X$ | The value of the measurable property MP is X |
| $QR = Y$ | The value of the quality requirement QR is Y. Y can be acceptable, tolerable, or inacceptable. |
| $InBindigs(SC)$ | Returns a list of all connectors bound to SC |
| $OutBindigs(SC)$ | Returns a list of connectors that SC is bound to them |

The components and connectors in the context model are a reflection of the running system. There is a causal connection between each component/connector and a model at runtime in the context. The component and connector model can be changed through the adaptation cycle. The changes only happen in the connectors and the bindings. Each change in the model (change a connector or a binding) is enacted in the running system by the BASBA container through predefined actions embedded in the connectors. BASBA container has the responsibility to add or remove connectors, and each connector has methods to change the bindings.

Measured properties and requirement satisfaction levels in the context model are updated by checkpoints. In order to do this, the data is collected by interceptors, and evaluated in the related checkpoints to calculate the value of measurable properties. If the value of a measured property shows a change in quality satisfaction level, the related quality value will also be updated in the context model. Inacceptable or tolerable quality satisfaction levels can start the adaptation planning unit to rectify the situation. The planning unit tries to detect a sequence of tactics with a desirable result, which can be executed in accordance with the state of the context model. Analyzing the result and possibility of executing tactics is done according to adaptation tactic templates.

Adaptation tactics are defined in first-order logic, providing the ability to apply changes. Each adaptation tactic consists of six parts: supporting connectors, precondition, pre-state, change action, post-state, and expected effect. The supporting connectors are units in BASBA that support enacting the tactic to the running system. The precondition is declared in first-order logic and determines whether the tactic can be executed on the basis of the context. The pre-state shows the state of the component and the connector before applying the tactic. The change action is a sequence of actions to apply the tactic. Actions might include "adding a connector", "removing a connector", or "changing a binding". Post-state shows the state of the component and the connector after applying the tactic. The pre-state and the post-state are simple propositions on components, connectors and bindings. The expected effect formulates the effects of executing the tactic on the measurable properties. For each measurable property affected by applying the tactic, a formula or a delegated function should be defined to model the effects.

Fig. 9 shows the parallel execution tactic as an example. In this example, applying the parallel execution tactic in component $SC$ is introduced. This tactic is supported in BASBA by *ParallelOutConnector* and *ParallelInConnector* connectors. *ParallelOutConnector* is a unit that simultaneously executes the next unites. *ParallelInConnector* is a unit that waits until it gets the first response, and then continues the process. The precondition shows that the tactic can only be applied if there is a component SC′ which is the same type as SC. The pre-state shows the components, connectors and the bindings before applying the tactic. Change action determines that for applying the tactic, a sequence of changes should be enacted. First, a connector of the type *ParallelOutConnector* and a connector of the type

*ParallelInConnector* are instantiated and added to the model. Next, these connectors are bound to SC and SC′. Then, each binding from and to SC is modified to the new added connectors. The post-state in the Fig. 9 shows the model of the system after applying the tactics.

The expected effect shows the effects of the tactic on measurable properties. In this example, we supposed a process block (pb) with three defined measurable properties: availability, cost, and response time. Due to the insignificant impact of connectors on measurable properties, the effects of connectors are ignored in the model.

Adaptation tactics in BASBA can be extended. For this purpose, a template class with the introduced parts should be defined for each tactic. In addition, if there is a need for a new connector that does not exist, the new connector should be implemented and added to BASBA connectors.

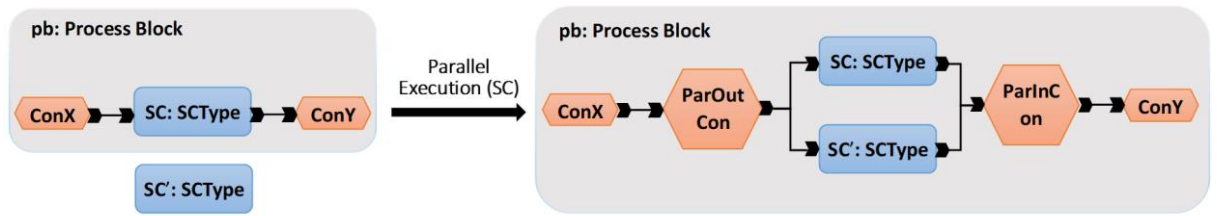

**Adaptation tactic:**
$ParallelExecution(SC)$

**Supporting connector:**
$ParallelOutConnector, ParallelInConnector$

**Precondition:**
$\exists SC': SC' \in ServiceComponent \land SCType(SC') = SCType(SC)$

**Pre-state:**
$SC: ServiceComponent, SC': ServiceComponent, ConX: Connector, ConY: Connector, bind(ConX, SC), bind(SC, ConY)$

**Change action:**
**add** $(ParOutCon: ParallelOutConnector)$
**add** $(ParInCon: ParallelInConnector)$
**add**$(bind(ParOutCon, SC)),$ **add**$(bind(ParOutCon, SC')),$ **add**$(bind(SC, ParInCon)),$ **add**$(bind(SC', ParInCon))$
**foreach**$(ConX: InBindigs(SC))$ {**remove**$(bind(ConX,SC)),$ **add**$(bind(ConX, ParOutCon))$}
**foreach**$(ConY: OutBindigs(SC))$ {**remove**$(bind(SC, ConY)),$ **add**$(bind(ParInCon, ConY))$}

**Post state:**
$SC: ServiceComponent, SC': ServiceComponent, ConX: Connector, ConY: Connector, ParOutCon: Connector,$
$ParInCon: Connector, bind(ParOutCon, SC) \land bind(ParOutCon, SC') \land bind(SC, ParInCon) \land bind(SC', ParInCon)$
$\forall ConX: bind(ConX, SC) \rightarrow bind(ConX, ParOutCon)$
$\forall ConY: bind(SC, ConY) \rightarrow bind(ParInCon, ConY)$

**Expected effect:**
$Availability(pb) = 1 - (1 - availability(SC)) * (1 - availability(SC'))$
$Cost(bp) = Cost(SC) + Cost(SC')$
$ResponseTime(bp) = Min(ResponseTime(SC) + ResponseTime(SC'))$

**Fig. 9** Parallel execution tactic at runtime

# 7. Evaluation

The general objective of the experimental study was to evaluate the effectiveness of BASBA framework to facilitate and improve developing adaptive behaviors in SBAs. We defined the experimental design of our study using the Goal-Question-Metric method [53]. The goal, questions, and metrics of the study, following the GQM template, are presented in Table 6.

Table 6 GQM template for BASBA framework evaluation

| | | |
|---|---|---|
| Goal | Purpose | Evaluation |
| | Issue | Effectiveness (impact on efficiency and quality of developed adaptive behaviors) |
| | Object (product) | BASBA framework |
| | Viewpoint | Development team |
| | Context | Service-based applications |
| Question | Q1 | Does BASBA improve identification of adaptation plans in the target system in comparison to conventional methods? |
| Metrics | M1 | The number of appropriate adaptation plans identified. |
| Question | Q2 | Does BASBA improve the efficiency of developing adaptive behaviors? |
| Metrics | M2 | Development time |
| Question | Q3 | Does BASBA increase the code quality of developed adaptive behaviors? |
| Metrics | M3 | The rate of faults/correctness of realized adaptation plans |
| | M4 | Understandability |
| | M5 | Modifiability |

These questions are designed to test the hypothesis that development of adaptive SBAs using BASBA is more efficient than traditional methods in terms of quality and development time. However, in addition to these questions, there are two other independent variables that could have a significant effect on the results: i) the professionality level of developers, and ii) the complexity level of adaptation needs and business processes. In the study, therefore, these parameters were also considered as independent variables and their effects were analyzed.

Regarding these questions, in order to evaluate the proposed framework, two case studies and an evaluation in academic environment were undertaken. In case study 1, we conducted a semi-controlled experiment in an industrial environment with full-time developers for two months. Case study 1 yielded interesting results; however, it was not enough for drawing a reliable conclusion about BASBA. It was difficult to replicate a similar study due to cost and resource limitations. To address the situation, we decided to analyze the effects of BASBA in one of the projects that have employed the BASBA framework in their development stack. Case study 2 was an exploratory research based on a flexible research design. The results of case study 2 showed that BASBA is an effective tool. However, because it was not a controlled experiment and the analysis was only based on a developed program and estimations, it was difficult to analyze the effects of independent variables on dependent variables. In order to have a complementary study, we conducted a controlled experiment in an academic environment using some scenarios from case studies 1 and 2.

### 7.1. Case study 1: Emergency and dispatching system

For the experiments, an emergency and dispatching system has been selected as the first case study. The motivation for choosing such a system is the problems encountered in emergency and dispatching scenarios. The system is a real case and has been operational for over one year. It is distributed throughout the entire country and receives over 40 thousand calls per day. It is operational in about 300 centers and runs on more than 5 thousand vehicle devices, which are connected through an unreliable private radio network. Architecturally, the case is a hybrid client-server/peer-to-peer system, in which the connection between clients and centers is based on client-server architecture, and the one between clients is based on peer-to-peer architecture.

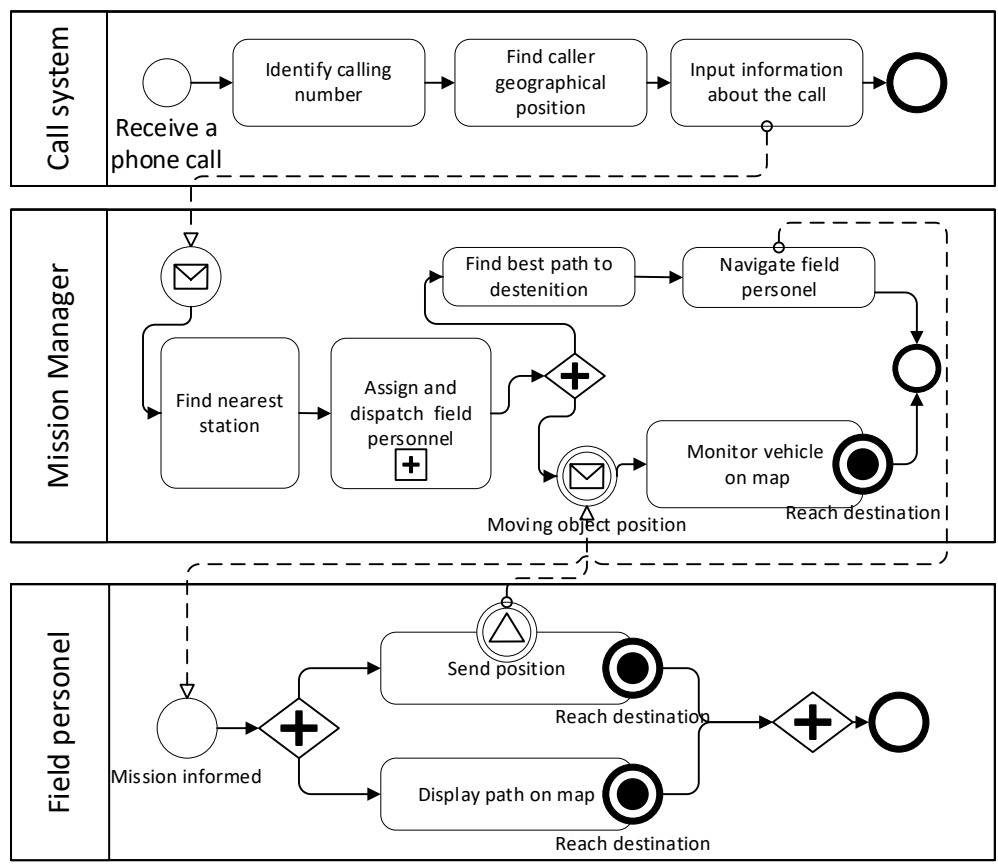

**Fig. 10** The process model of the emergency system

Fig. 10 shows the overall model of one process of this case study. For the sake of simplicity, many elements and other related workflows are left out of this paper. Using the system's log, over 80 scenarios of contextual and environmental changes were identified. These scenarios were considered as a reference for determining adaptation needs, and were also simulated to evaluate the developed adaptive behaviors through the experiments. It should be mentioned that all of these scenarios were handled by the running emergency system, which provides a proper reference for comparison. Nevertheless, they were all developed using conventional methods, mostly with fixed hardcoded alternatives.

The usual concern of the case is to deal with different environmental contextual changes such as unreliability of network or unavailability of services. The system should perform self-adaptive behaviors in order to maintain three quality objectives: i) reliability, ii) availability, and iii) response time.

A group of 11 full-time software engineers participated in this experiment for about two months. All the participants were familiar with the case study and were members of the company where the real case was developed. They included a project manager, an architect, an analyst, and eight developers. The project manager, the architect and the analyst were members of a team who had been responsible for developing and maintaining the real case. In the study, they were experts who observed, controlled, and analyzed the results.

Regarding the effect of developers' professionality, the developers were divided into two groups. The first group consisted of four nonprofessional developers who had 2-4 years of experience in programming without practical knowledge about how a system should be adapted to contextual and environmental changes. The second group was made up of four developers who had at least eight years of experience in programming with deep knowledge about design patterns. They were also familiar with developing adaptive behaviors, and each of them had at least the experience of developing one complex system requiring adaptive behaviors. We divided each group into two subgroups: one was to develop adaptive behaviors using BASBA and the other to do it without using BASBA, relying on their personal expertise and experience. The reason for this subdivision was to specify the contribution of BASBA to the development of adaptive behaviors in different groups with different expertise.

Regarding Q1, we asked each group to identify adaptation plans to handle possible environmental and contextual events. Furthermore, we asked the project manager, the architect and the analyst to analyze and categorize all the defined scenarios and determine all the adaptation needs. Regarding the effect of complexity variable, we asked them to rank the complexity of the adaptation plan from adaptation and business perspectives from A (the lowest degree of complexity) to C (the highest degree of complexity). We then asked the experts to analyze the results of each group to determine the extent to which each group was successful in identifying these adaptation needs.

**Table 7** The role of BASBA in identifying proper runtime adaptation tactics for ten adaptation plans in detail

| Adaptation plan | Complexity | Adaptation category | Adaptation tactics reused | Nonprofessionals without BASBA | Nonprofessionals using BASBA | Professionals without BASBA | Professionals using BASBA |
|---|---|---|---|---|---|---|---|
| Replace automatic call detection with manual | AA | process | replace to manual | ✔ | ✔ | ✔ | ✔ |
| Skip automatic call detection, Find caller geographical position on map | BB | process | skip, add activity | ✘ | ✔ | ✔ | ✔ |
| Skip automatic call detection, Input caller text address, Geocode text address | BC | process | skip, add activity, add activity | ✘ | ✘ | ✔ | ✔ |
| Automatic call detection, Input caller text address, Manual station selection | CB | process | skip, add activity, replace to manual | ✘ | ✔ | ✘ | ✔ |
| Serial execution of map by invoking from Municipality, Google and Bing | BA | activity | serial execution | ✘ | ✔ | ✔ | ✔ |
| Re-execution of asking to dispatch personnel | AB | activity | re-execution | ✔ | ✔ | ✔ | ✔ |
| Insert compressor and decompressor between send and receive vehicle data | CA | communication | insert compressor and decompressor | ✘ | ✔ | ✘ | ✔ |
| Add queue before sending message to vehicles | CB | communication | add queue | ✘ | ✔ | ✘ | ✔ |
| Add local cache for map | BB | communication | local cache | ✘ | ✔ | ✔ | ✔ |
| Display vehicle place with text address | BC | communication | reduce size | ✘ | ✘ | ✔ | ✔ |
| Transform points to trajectory | CC | communication | aggregate data | ✘ | ✘ | ✘ | ✘ |

The results of the experiment for metric M1 are described in Tables 7 and 8. In these tables, adaptation and business complexity are shown with two characters, the first of which shows adaptation complexity and the second one shows business complexity. Table 7 shows the results of 10 adaptation plans in detail. For each adaptation plan the complexity, the adaptation category, the adaptation tactic, and the groups that managed to identify the required adaptation plan are determined. Table 8 summarizes the results for all the 24 adaptation plans. It shows the number of adaptation plans in each category. For each group, the cell shows the number of adaptation plans identified by them. The identified plans are separated by semicolons. The table also shows the comparison between "group 1, using BASBA" and "group 2, without BASBA". Each bold-faced item shows a plan identified by a group and not identified by the other group. The purpose of the comparison was to analyze the extent to which BASBA can enhance nonprofessional developers' ability to identify adaptive behaviors in comparison with professionals. As shown in the tables, there is a tangible improvement in identifying adaptive plans (metric M1) where BASBA is employed. While nonprofessional developers without BASBA managed to identify 6 appropriate adaptation plans, the number increased to 14 when the BASBA framework was employed. For the professional users, the number was increased from 14 to 20. The results show that plans with low adaptation complexity were identified by almost all the groups. However, BASBA was very useful with plans with high adaptability complexity and low or medium business complexity. Nevertheless, it was not very effective in dealing with plans with high business complexity. The results show that there was a tangible enhancement in identifying adaptation tactics in all three categories of adaptation tactics. Particularly, there was a significant enhancement in identifying communication tactics.

**Table 8** The summarized results of all 24 adaptation plans

| Adaptation category | Total adaptation plans | Nonprofessionals without BASBA | Nonprofessionals using BASBA | Professionals without BASBA | Professionals using BASBA | Not detected plans |
|---|---|---|---|---|---|---|
| Process variation | 10 | AA; AB; AB | AA; AB; AB; BA; BB; BB; **CB** | AA; AB; AB; BA; BB; BB; **BC** | AA; AB; AB; BA; BB; BB; BC; CB; CB | BC |
| Activity variation | 6 | AA; AB; BA | AA; AB; BA; BA | AA; AB; BA; BA; **BB** | AA; AB; BA; BA; BB; BB | - |
| Communication variation | 8 | - | BB; **CA**; **CB** | BB; **BC** | BB; CA; BC; CB; CB | BC; CC; CC |

Regarding Q2 and Q3, we asked each fully-dedicated (eight hours per day) group to develop all the 24 defined adaptation plans, and then their performance was evaluated in terms of development time and code quality. Regarding metric M2 in Q2, the development time was measured by total working days. The results show that nonprofessional developers (not using BASBA) managed to develop only 12 adaptation plans in 26 working days, while the nonprofessional developers who used BASBA managed to develop 20 adaptation plans within almost the same span of time which included all the 12 adaptation plans developed by the group without BASBA. The average development time was reduced from 2.17 to 1.3 working days per adaptation plan, that is, about 40 percent saving in time. The average development time for the same 12 adaptation plans developed by both groups reduced from 2.17 to 1.25 working days per adaptation plan. The time records showed that the same 12 adaptation plans were developed almost in 15 days by the second group. Both teams in group 2 (professionals) managed to develop all the adaptation plans. However, those who used BASBA did so in 15 working days, and those who did not use BASBA managed to develop the adaptation plans in 21 working days. The average development time for the professionals was reduced from 0.88 to 0.63 working days per adaptation plan, that is, about 29 percent saving in time.

Regarding metric M3 in Q3, the correctness of code was measured by the number of faults emerged through evaluation of the developed adaptation plans in test scenarios. The results show that ordinary developers (not using BASBA) recorded 57 faults in 12 adaptation plans (on average, 4.7 faults per adaptation plan), while ordinary developers who used BASBA recorded only 52 faults in 20 adaptation plans (on average, 2.6 faults per adaptation plan), which means the average number of faults per adaptation plan was decreased for about 45 percent. Regarding the 12 adaptation plans which were the same as the ones developed by the ordinary developers (not using BASBA), the number of faults was 28 (on average, 2.3 faults per adaptation plan) meaning an over 50 percent improvement. The result for professionals who had not used BASBA was 46 faults in 24 adaptation plans (1.9 faults per adaptation plan on average), and for professionals who used BASBA it was 31 faults in 24 adaptation plans (1.3 faults per adaptation plan on average), which means an improvement about 32 percent.

Regarding the role of BASBA in understandability (M4 in Q3) and modifiability (M5 in Q3), the developed adaptation plans were analyzed by the experts in focus group sessions. The group members (the project manager, the architect, the analyst, and the moderator) agreed that the developed adaptation plans with BASBA were more understandable and modifiable.

During the experiments, shadow observations were carried out by the experts, who observed the behaviors of each group and noted their observations. The notes were coded and categorized to quantify qualitative data. The high-frequency codes, related to question 2, were selected and analyzed. The results are summarized as follows (*Italic* expressions are frequent codes):

BASBA facilitated the development of adaptive behaviors by *reducing the complexity* of the problem through *breaking it down into simpler sub-problems* and *separate adaptation logics*, resulting in a more *understandable and maintainable code* (metrics M4 and M5). Furthermore, it helped to *easily test* the developed adaptive behaviors, and improved control over the adaptive behavior by enhancing *traceability* and providing the ability to *check the overall outcome of the adaptation plans*, which resulted in improving the *correctness of code* (metric M3). Moreover, BASBA enabled developers to *reuse adaptation tactics* and implement the adaptation logic by *less lines of code* that resulted in *reducing the development time* (metric M2).

### 7.2. Case study 2: Regional power distribution management system

For the second case study, a regional power distribution management system (RPDMS) was selected. The motivation for choosing this system was that the development managing board of RPDMS had decided to use BASBA framework in the development process. The system had been planned to be employed in operation, monitoring, and maintenance of the power distribution infrastructure in a state. The infrastructure includes a few thousand sensors and devices maintained by over 500 staffs in different teams. These teams, which include management, monitoring, and operational teams, were equipped with portable devices such as PDAs and car PCs to have real time communication. The system operates over different areas with unpredictable environmental conditions, such as unreliable network. However, the system should behave in dependable manner and perform self-adaptive behaviors. For example, as a team member moves away from a WLAN coverage, the PDA loses connection to the local location service. In this situation, several conditions may happen. If the connection switches to GPRS by the software installed on the PDA, the location service can be obtained from the GPRS provider with a low accuracy. However, the team member may get into the car equipped with a navigation system based on GPS, which provides a more accurate location service. To increase accuracy, the application can reconfigure itself to use location services provided by car devices. To save battery life, the display can also switch to the car display system.

In another example, in some areas there may not be any internet connection. In this situation, it is very important to maintain interaction between team members. One option is to maintain communication via SMS over GSM network. However, in this situation, only important parts of messages without any additional data may be transferred and the full messages should be queued until reaching an area with internet connection. Compressing SMS messages will also be another option if there is enough battery life.

In order to develop the system, a team with 17 members were assigned to the project. The development methodology was SCRUM. At the start of the project, the BASBA framework was introduced and explained to the team members. Through the development process, the team members had access to BASBA experts whenever they needed some help. The system was developed in 8 months. At the end of the project, the effectiveness of BASBA was analyzed. Overall, 34 considerable adaptation scenarios were implemented. All of these adaptation scenarios were analyzed by the development team members and BASBA owners.

Regarding the effect of developers' professionality, we prepared a questionnaire for the development team members. The questionnaire included two main categories: i) How much programming expertise is needed to learn the BSABA framework, ii) How much adaptation skill is needed to develop adaptation strategies using BASBA. Regarding the former question, all the participants agreed that both senior and junior developers could learn BASBA. As for the latter question, 88 percent of participants (15 out of 17) stated that basic knowledge of developing adaptive behaviors is enough to develop adaptation strategies using BASBA.

Regarding metric M1 in Q1, we analyzed the log activity of identifying adaptation strategies, and asked the developers to determine the role of BASBA in the identification process. Plans were considered as "identified by BASBA" when the majority of the development team (over 70 percent) believed that BASBA had a significant role in the identification process. The results show that 13 (out of 34) adaptation plans were identified using BASBA framework (Table 9).

Regarding the complexity variable, in Table 9, each scenario is shown with two characters, the first of which shows adaptation complexity and the second one shows business complexity (from A to C). The results show that BASBA had a more considerable role in identifying activity and communication plans (7 plans out of 12) with low or medium business complexity. BASBA was not helpful in any plan with high business complexity.

**Table 9** The summarized results of all 34 adaptation plans identified

|  | Number of process variations | Process variations | Number of activity variations | Activity variations | Number of communication variations | Communication variations |
|---|---|---|---|---|---|---|
| Identified using BASBA | 3 | BA; BB; BB | 3 | BA; BB; CB | 7 | BA; BA; BB; BB; CA; CA; CA |
| Not identified using BASBA | 11 | AA; AA; AA; AB; AB; BB; AC; AC; AC; BC; BC | 5 | AA; BB; AB; AB; BC | 5 | AC; BB; BC; CC; CC |

In the project, 32 adaptation plans were developed using BASBA framework. These plans included 14 process variations, 8 activity variations, and 10 communication variations. Two communication adaptation plans were too complicated and the architect of the system decided not to implement them using the BASBA framework. For example, implementing "skip communication with servers and switch to P2P communication" was implemented without BASBA. Table 10 shows some adaptation plans implemented using BASBA.

**Table 10** Some developed adaptation plans in RPDMS

| Adaptation category | Adaptation plan |
|---|---|
| Process | • Skip monitoring team members<br>• Switch from map visualization to text display<br>• Replace push notification with pull notification<br>• Get location service from car<br>• Replace WLAN with GSM |
| Activity | • Decrease/Increase monitoring interval<br>• Send data until receive acknowledge<br>• Re-execution asking to dispatch personnel |
| Communication | • Skip encrypting and decrypting data<br>• Add queue before sending messages<br>• Send data in urgent mode<br>• Send only essential data |

Regarding Q2 and Q3 (the role of BASBA in automating and facilitating the development of adaptation plans), a questionnaire was prepared to evaluate development time and code quality metrics, and the development team were asked to analyze the development log and response to the questionnaire. Each plan was analyzed by at least two developers on the basis of their previous experiences and developed projects log. The development team estimated that the development time of implementing adaptation plans had decreased for all of the developed plans by at least 35 percent on average (metric M2). In terms of code quality, the analysis showed that the code of adaptation logic in almost all developed plans had become more readable and more structured (metric M4) in comparison with previously developed projects. The development team did not come to a clear conclusion about the role of BASBA on metrics M3 and M5.

### 7.3. Evaluation in an academic-environment study

In order to have a deeper analysis about the role of BASBA framework in developing adaptation plans, we conducted a supervised evaluation in an academic environment. For this purpose, two groups took part in the experiment. The first group included 17 professional software engineers from Service-Oriented Enterprise Architecture Laboratory[1] (SOEAlab), and the second group consisted of 38 M. A. students of software engineering, most of whom were junior software developers. The experimental material was composed of 42 scenarios selected from case 1 and case 2. The scenarios were categorized in three levels of complexity. In the first step of evaluation, we selected fifteen scenarios including five random scenarios from each category for each subject. The participants were divided into traditional and BASBA groups. The traditional group included 18 students and 7 members of SOEAlab who were responsible for developing adaptation plans by means of traditional solutions. The BASBA group included 20 students and 10 SOEAlab members to whom BASBA framework was introduced. The subjects were asked to identify a proper adaptation plan for the scenarios and write pseudo-codes to show how they implement the plans in 150 minutes. The objective of these steps was to evaluate the role of BASBA in identification and development of adaptation strategies. The results of the experiments were analyzed by five experts including two developers from case study 2 and shown in Fig 10, 11 and 12. We used boxplot charts to summarize the results, and two-tailed, paired t-tests to evaluate the statistical significance of the results.

Regarding Q1, the experts analyzed the results to measure if the subject identified a proper adaptation plan for each given scenario. The experts gave a score between 0 to 10 to show the quality (properness) of the identified and realized adaptation plans. The number and quality of identified adaptation plans for each group are shown in Fig. 11 (a) and (b). The results show that the number and quality of identified adaptation plans present a significant statistical difference in comparison to traditional methods. The average number of identified plans increased from 6.3 to 10.3 for students, and from 11.4 to 13.6 for SOEAlab members. Regarding the quality (appropriateness) of identified plans, the figures increased from 3.5 to 5.8, and from 5.1 to 8.2 respectively. The results show that students who used BASBA have identified more appropriate adaptation plans than those who have not used BASBA (p-value <0.01). They performed the tasks in almost the same manner as the experts without BASBA. The figures also show that BASBA had a tangible effect (p-value <0.05) on identifying adaptation plans for the experts.

---

[1] https://soea.sbu.ac.ir/en

Regarding metrics M2 and M3, the experts analyzed the number and correctness of realized adaptation plans. In this study, metric M2 (development time) was measure by the number of realized adaptation plans in the specified time frame (150 minutes). The experts gave a score between 0 to 10 to show the correctness of the realized adaptation plans. The number and correctness score of realized adaptation plans for each group are shown in Fig. 12 (a) and (b). The results show that the number and correctness of realized adaptation plans present a significant statistical difference in comparison to traditional methods. The average number of realized plans increased from 5.5 to 7.9 for students, and from 8.7 to 10.6 for SOEAlab members. Regarding the correctness of realized plans, the figures increased from 2.3 to 5.5, and from 4.3 to 7.5 respectively. The results show that employing BASBA had a significant effect on the number and correctness of realized adaptation plans for the students, and on correctness of realized adaptation plans for the experts (p-value <0.01). The statistical difference of correctness for the experts was also tangible (p-value <0.05).

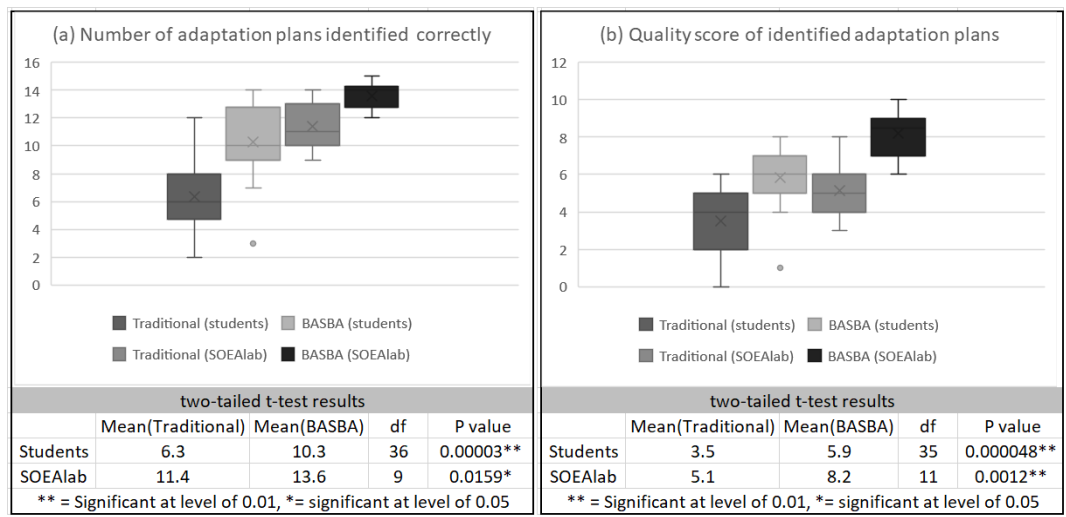

**Fig. 11** Boxplot graphs and t-test results for the number and quality of identified adaptation plans

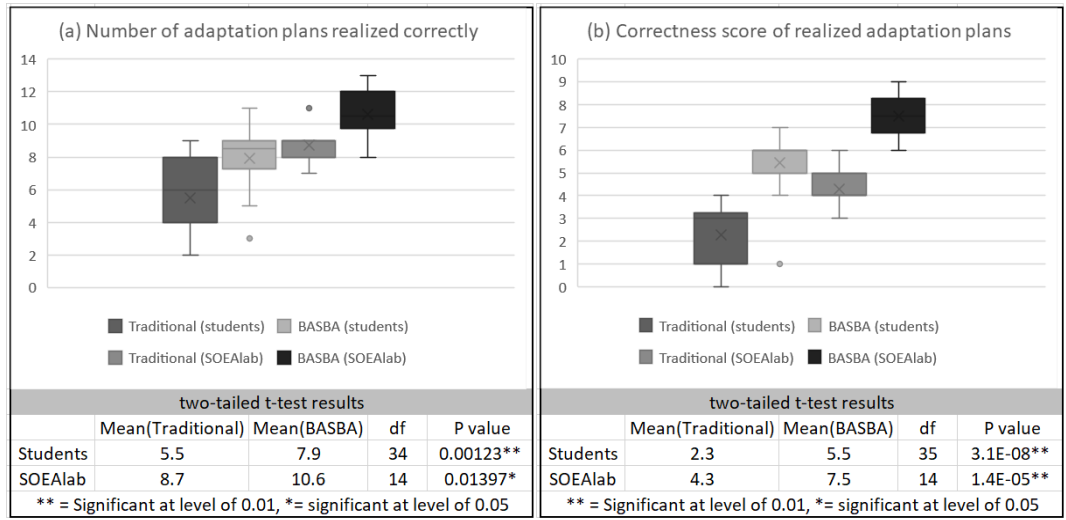

**Fig. 12** Boxplot graphs and t-test results for number and correctness of realized adaptation plans

Regarding metrics M4 and M5 (understandability and modifiability) in Q3, the codes of 12 adaptation scenarios with different structural complexities were selected. Each subject was given all scenarios, but 6 random scenarios with traditional implementation and the remaining scenarios with BASBA implementation. For each scenario, two questions were asked: one of which was to answer the understandability about the logic of adaptation, and the other containing a modification task about changing the adaptation logic and how the change should be enacted in code. In the experiment, each subject answered all the 24 questions (two questions for each given scenario) in 150 minutes. The answers were analyzed by the experts and received feedback on whether the subject had understood the adaptation logic, and whether he/she was successful in enacting the changes. The average scores of understandability and modifiability, given by the experts, were mapped to a number between 0 to 10. The boxplot charts and t-test values of the results are represented in Fig. 13 (a) and (b). As the results show, the understandability is increased in average from 3.8 to 7.5 for students, and from 6.7 to 8.9 for SOEAlab members, indicating a significant statistical difference for plans implemented with the BASBA framework. Regarding modifiability, the figures increased from 3.1 to 4.2 for the students, showing a significant difference; for the experts, however, the difference between means was not statistically significant (p-value = 0.052> 0.05).

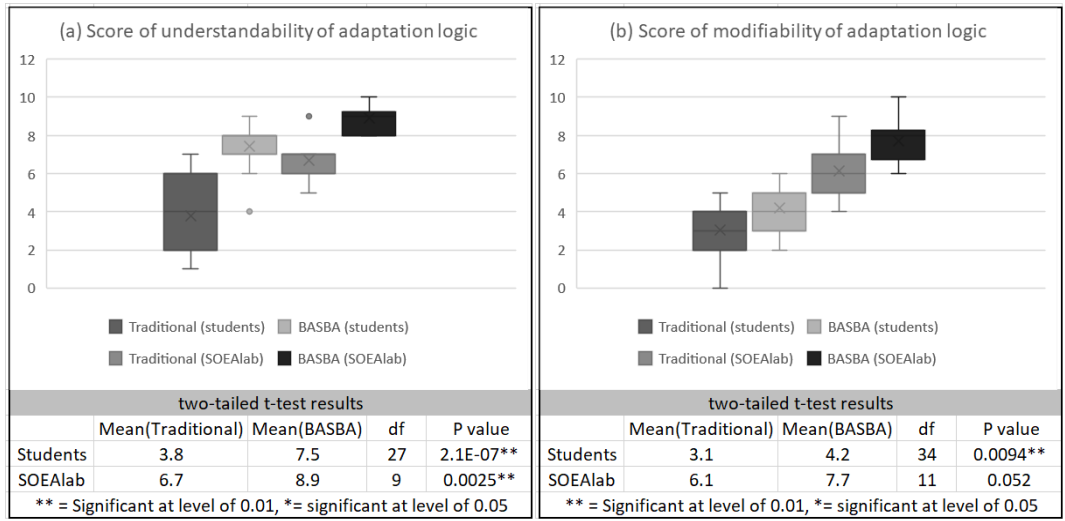

**Fig. 13** Boxplot graphs and t-test results for understandability and modifiability

### 7.4. Threats to validity

Empirical evaluation is always subject to different threats that can influence the validity of the results. It is not always possible to address all the threats in one study. To mitigate this, we conducted three studies to evaluate BASBA, which can complement each other and improve the validity of the study results. In case study 1, different aspects of BASBA were investigated in a semi-controlled experiment. However, the limited number of subjects and the difficulty to have a replicate study pose different threats to the validity of the results. In case study 2, the effectiveness of BASBA in a real project was investigated. However, in the experiment there was no proper control on independent variables. In addition, the history of the company and estimations of measure parameters can threaten the validity of results. In study 3, a meaningful number of subjects participated in the experiment. However, different factors, such as the limitation on time, unfamiliarity of participants with the study and learning effects during the training session can threaten the validity of results. Here, we will discuss the threats to conclusion, construct, internal, and external validity in more detail. Our goals are to help readers qualify the results and to highlight the aspects of our experiments that may have been affected by these threats.

**Conclusion validity:** Conclusion validity concerns the relationship between the treatment and the outcome. One threat, here, is about statistical validity of the experiments. The number of developers in case study 1, and analyzing only one project in case study 2, could affect the statistical validity of the results, which can cause limitation to drive the conclusion. To mitigate this in study 3, a meaningful number of participants were involved. Another threat is related to misinterpretation of the results, particularly when it comes to some qualitative interpretations such as code quality. To mitigate the threat in case study 1, we tried to use quantitative measurements such as the number of faults and working days. In addition, the results of all cases were interpreted by experts with enough knowledge in the domain. Another threat is fishing for the result. To mitigate this in case study 1, there was not any advantage in the results for the participants. In addition, the developers regard the experiment as a part of their regular work, and they were not aware about the measured parameters. However, the motivation to learn a new concept, particularly for the nonprofessional developers, can be an intervene factor, which may have been ignored. In studies 2 and 3, this issue is aggravated, as most of the experts were from the company where BASBA was applied. To mitigate this, each result was evaluated by 3 to 5 experts. In addition, in study 3, some experts were not from the developers' company. Another issue is reliability of the measures. Measurement of different properties, such as quality and time, in developing adaptive behaviors can be distorted by different variables such as prior experience or environmental factors. To mitigate this, in case study 1, we tried to maintain the same environmental factors for all the developers with almost the same working hours. In addition, the activities of developers were precisely observed and analyzed through shadow observation to verify the results and detect the uncontrolled factors. These arrangements combined with the fact that there is no gain for a participant in adjusting their measurements, the reliability of measures should be good in case study 1. Moreover, in case study 2, we analyzed the outcome of BASBA in a real project and in study 3, an experiment with a meaningful number of subjects was conducted.

**Internal validity:** Internal validity is the extent to which a piece of evidence supports a claim about cause and effect. One threat, here, is ignoring relevant factors. To mitigate this, in case study 1, shadow observation was employed during the experiment to analyze and resolve the impact of other independent variables. In study 3, we conducted the experiment in a short period of time with specific questions. However, in case study 2, because the analysis was done after developing the system, we were not able to analyze the effects of other independent variables. Another issue is learning effect in studies 1 and 3. To mitigate this, in case study 1, we used different groups to have the same learning effect. In study 3, a random selection and order of scenarios was used to avoid learning effects. However, in part 1 of this study, one group had a training session, which could have an undesirable side-effect on the results and threaten the validity. Another issue is subject selection. Case studies 1 and 2 were conducted in a single company that concentrates on developing location-based systems. This limitation could threaten the validity of the results, because the subjects were not heterogenous. In case study 1, we tried to have a slight mitigation by involving different roles with different expertise. In addition, we conducted study 3 as a complementary study with heterogeneous subjects from academic environment.

**Construct validity:** Construct validity refers to the belief that the dependent and independent variables represent the theoretical concept of the phenomenon accurately. One threat here is the definition of metrics for code quality. In the first experiment, the metric for measuring correctness of realized adaptation plans was defined on the basis of the number of faults in the code. In addition, understandability and modifiability were analyzed by the experts. However, in case study 2, we only managed to measure understandability of the code. Another threat is the measurement method and the details of the measurement, which affect the study results. We know that the complexity of measuring qualitative metrics could produce false estimations and threaten the validity of the results. To mitigate this in case study 1, we tried to define quantitative metrics. However, in studies 2 and 3, we used estimations to analyze the results whose deviation in estimations can threaten the validity of the results. Particularly, when it comes to measuring code quality objective, the matter gets worse because of the lack of a proper mechanism to estimate the role of BASBA in improving the objective. To mitigate this threat, several experts with proper knowledge are involved to estimate the results. However, we think more studies like case study 1 need to be done. Another threat is experimenter bias. Particularly, in studies 2 and 3 the metrics were measured by the experts based on their opinion. The threat posed by using expert measurement mechanisms is that different experts may have different attitudes toward the evaluation of dependent variables. For instance, some experts may be reluctant to use some kind of adaptation mechanism or code style, or provide a solution different from BASBA. To mitigate this, in case study 1, the experts controlled this by shadow observation, and in study 3, three experts not involved in case study 2 were asked to contribute.

**External validity:** External validity is the extent to which the study can be generalized to other subject populations and settings. Regarding the external validity, threats originate from how the study can be generalized to other subject populations and settings. To limit this threat, external developers from the field are involved to increase the relevance of the study to real applications. Furthermore, developer activities were analyzed through shadow observation in case 1, and the log of activities were analyzed in case 2. This analysis shows that the scenarios can be applicable in other similar cases. In addition, we conducted an academic environment study to analyze the role of BASBA to facilitate developing adaptation plans. However, we cannot claim that the same result can be obtained for any system or any situation. Especially, since all the experiments were conducted by one company and the scenarios were biased toward the domain of location-based services. In fact, in all cases, most of the adaptation plans was related to location. However, applying BASBA in location-based systems shows promising results, but there could be doubts on the applicability of BASBA in other domains. We believe that BASBA should be evaluated in other industrial environments, particularly those which are not location-based.

We are aware that these issues may pose threats to the reached conclusions, so the results of these experiments were considered as preliminary findings. We are currently working on some other projects that are implemented by BASBA to conduct replication studies.

### 7.5. Discussion

In this section, we reflect upon our experiences with applying the BASBA framework. Our remarks are based on the authors' experience of applying BASBA to the three experiments explained previously, as well as discussions with several developers from the company where the BASBA framework was developed and employed. We present a number of the benefits of the BASBA framework, and also outline some limitations and research challenges with regard to how/when to use the BASBA framework for the development of adaptive behaviors in SBAs.

The BASBA framework provides us with a means to identify and develop some repetitive adaptive behaviors based on reusing defined adaptation tactics. In this regard, BASBA introduces a systematic approach to develop adaptive behaviors based on specific engineering of adaptation engines and feedback loops, with the possibility to reuse adaptation tactics. The results of the experiments show that employing the BASBA framework can result in identifying more appropriate adaptation plans as well as enhancing the development efficiency (development time) and quality (correctness, understandability, and modifiability) of developed adaptive behaviors. It helps to keep adaptation concerns and behaviors at the design level separate from execution models, reducing the complexity and increasing the maintainability of the system.

While BASBA shows promise as a framework for developing adaptive behaviors in SBAs, there are still several issues related to some aspects of applying BASBA. We have identified adaptation tactics as a desirable capability for developing adaptive behaviors, but the capability of BSABA is limited to the developed adaptation tactics. BASBA has provided the ability to extend adaptation tactics. However, when an adaptation needs is complex, particularly in terms of business process, or it is strongly domain specific (like some scenarios in case study 2), the BASBA framework does not seem very effective.

BASBA was an effective tool for both professional and nonprofessional developers in all the cases, however, the results show that professional developers did not enjoy its benefits as much as nonprofessional developers. Particularly, for plans with business complexity, the professional developers did not find BASBA a useful tool. The results show that although BASBA was an effective means to identify plans with high adaptability complexity and low or medium business complexity, it was not very effective in dealing with plans with high business complexity. This issue was quite obvious in case study 1, where there was not a significant enhancement in identifying adaptation plans with high business complexity.

Another current limitation is that BASBA only addresses implementing adaptation behaviors based on defining BASBA adaptive process models, whereas in some cases, it is not efficient to define adaptive behaviors using the BASBA framework. The limitation was observable in case study 2, where the developers avoided implementing two adaptation plans based on BASBA adaptive process model. Besides, while the results show that BASBA had a significant role in improving the identification of adaption plans, it cannot be considered a comprehensive tool for identifying all adaptation plans. This issue became evident in case study 2, where the developers stated that BASBA had a significant role in identifying only 13 (out of 34) adaptation plans.

## 8. Conclusions

In this paper, we have presented a new approach to developing adaptive SBAs based on runtime models that allows for the specification and execution of adaptive processes. The core of this approach is a metamodel used to define adaptation behaviors in SBAs and a set of reusable adaptation tactics. On the basis of this metamodel and reusable adaptation tactics, a service integrator would be able to effectively identify and define adaptive behaviors for an SBA. The defined adaptive behaviors can then be efficiently transformed into runtime models to form adaptation logics. The proposed approach is automatically supported by feedback loops and runtime models. It noticeably facilitates the development of adaptive behaviors of an SBA by managing variations using runtime models, and execution of the adaptation engine.

In order to evaluate the proposed approach, we have conducted two real case studies and an academic-environment study. The results showed that using BASBA can effectively enhance the development process of adaptive behaviors in terms of development time and code quality, particularly for scenarios that are more complex in terms of adaptation rather than business logic. The introduced approach provides service integrators with the ability to deal with adaptation concerns in a more abstract way, relieving the service integrator of low-level monitoring and adaptation mechanisms.

There are several directions for future work. First, the adaptation tactics supported by BASBA framework need to be extended. Particularly, there should be tactics to support decentralized adaptation of service coordination. In addition, the monitoring mechanism in BASBA is limited to basic structures and events and a complex event processing can enhance BASBA significantly. Furthermore, a reusable model for analyzing the tradeoff among quality attributes based on different applicable adaptation plans can help improve the runtime decision of whether or not an adaptation plan should be applied. Moreover, the pattern-based analysis of adaptation plans can help reveal undesirable consequences associated with such plans. Finally, lessons can be learned from the execution and adaptation history to automatically improve the accuracy of the analysis.

**Acknowledgment**

The authors thankfully acknowledge the helpful support of "Knowledge City ICT Development Compony" in implementing and evaluating the BASBA framework under "Parsimap Intelligence" project (www.parsimap.ir).